\documentclass[12pt]{article}
\usepackage[english,activeacute]{babel}
\usepackage{natbib}
\usepackage{comment}
\usepackage{float}
\usepackage[hidelinks]{hyperref}
\usepackage{mathrsfs}
\usepackage{enumitem}
\usepackage[font={small,it}]{caption}
\usepackage{amsmath,amsfonts,amsthm,amssymb}
\usepackage{bm,rotating,multirow,dsfont,graphicx}
\usepackage[usenames, dvipsnames]{color}
\usepackage{url}
\usepackage{multicol}
\usepackage{multirow}
\usepackage[T1]{fontenc}
\usepackage{flafter}
\usepackage{appendix}
\usepackage{subfigure}
\usepackage{xcolor}
\usepackage{soul}
\usepackage{listings}
\usepackage{setspace}
\makeatletter
\def\hlinewd#1{%
	\noalign{\ifnum0=`}\fi\hrule \@height #1 %
	\futurelet\reserved@a\@xhline}
\makeatother
\usepackage[all,cmtip]{xy}

\def\exp#1{{\rm exp}{#1}}
\def\frac#1#2{{{#1}\over{#2}}}

\def\@roman#1{\romannumeral #1}

\begin{document}

\title{Parapolitics and Roll-Call Voting in Colombia: \\ A Bayesian Euclidean and Spherical Spatial Analysis}

\date{}
	
\author{
    Juan Sosa, Universidad Nacional de Colombia, Colombia\footnote{Corresponding author: jcsosam@unal.edu.co} \\
    Carolina Luque, Universidad Ean, Colombia \\
    Juan Valero, Universidad Nacional de Colombia, Colombia
}	 
        
\maketitle

\begin{abstract} 
This study presents a Bayesian spatial voting analysis of the Colombian Senate during the 2006–2010 legislative period, leveraging a newly constructed roll-call dataset comprising 147 senators and 136 plenary votes. We estimate legislators’ ideal points under two alternative geometric frameworks: a traditional Euclidean model and a circular model that embeds preferences on the unit circle. Both models are implemented using Markov Chain Monte Carlo methods, with the circular specification capturing geodesic distances and von Mises-distributed latent traits. The results reveal a latent structure in voting behavior best characterized not by a conventional left-right ideological continuum but by an opposition–non-opposition alignment. Using Bayesian logistic regression, we further investigate the association between senators’ ideal points and their involvement in the para-politics scandal. Findings indicate a significant and robust relationship between political alignment and para-politics implication, suggesting that extralegal influence was systematically related to senators’ legislative behavior during this period.
\end{abstract}


\section{Introduction}

Spatial voting models provide a powerful framework for analyzing legislative behavior by positing that political actors make decisions based on their proximity to ideological alternatives within a latent policy space \citep{luque2024operationalizing}. The central object of interest in these models is the \textit{ideal point}, a latent trait representing an individual’s political preference, estimated from observed voting behavior across a set of motions. These models have been extensively applied in the analysis of parliaments, senates, courts, and councils across diverse institutional contexts \citep{clinton2004statistical, poole2005spatial, curini2020sage}. Among them, Bayesian formulations offer a flexible and probabilistically coherent approach for ideal point estimation, allowing for principled uncertainty quantification and the incorporation of prior information \citep{jackman2001multidimensional, clinton2004statistical, luque2023bayesian}.

In this study, we advance the application of spatial voting models by conducting a detailed Bayesian analysis of roll call data from the Colombian Senate during the 2006–2010 legislative period, a term marked by profound political realignment and the emergence of the \textit{parapolitics scandal}, which revealed ties between legislators and illegal paramilitary groups \citep{avila2012politics, bakiner2020endogenous, pardo2020dual}. Building upon the methodology of \citet{luque2023bayesian}, who applied Bayesian spatial models to the Colombian Senate for the 2010–2014 period, we extend the analysis in both statistical and substantive dimensions. From a political science perspective, we provide one of the first systematic mappings of latent political preferences during a critical period of institutional transformation in Colombia. From a statistical perspective, we compare two distinct geometric formulations of spatial voting models: The traditional Euclidean model and its circular generalization, the Spherical Latent Factor Model proposed by \citet{yu2020spherical}.

A key contribution of this study lies in the construction of a comprehensive roll call database comprising 147 senators and 136 plenary votes from the 2006–2010 legislative period, enriched with detailed information on political affiliations, replacements, and each legislator's involvement in the parapolitics scandal. Using this dataset, we implement a one-dimensional Bayesian ideal point model under two distinct geometric frameworks. The first assumes a linear ideological continuum embedded in Euclidean space and is estimated via a probit formulation with normal priors. The second framework replaces Euclidean with geodesic distances by embedding ideal points on the unit circle, modeling latent positions with von Mises priors. This circular formulation addresses limitations of the Euclidean model by capturing potential behavioral convergence among ideological extremes and correcting distortions that arise from projecting complex political behavior onto a linear space \citep{tversky196919intransitivity, spirling2007uk}.

To ensure comparability across geometries, we align and project circular estimates onto the tangent space at a fixed reference point, enabling direct comparison with Euclidean ideal points. We also address model identifiability through anchoring and post-processing procedures, and handle missing voting data via complete case analysis in the Euclidean model and posterior imputation in the circular model, verifying the robustness of results across both strategies.

The estimated ideal points reveal a latent structure not reducible to a conventional left–right ideology. Rather, both geometric formulations uncover a broad opposition–non-opposition divide, with opposition, independent, and minority senators clustering separately from the dominant governing coalition. This non-ideological polarization reflects structural imbalances in the Senate and aligns with findings from prior work on Latin American legislatures \citep{londono2014efectos, luque2023bayesian}. Moreover, we explore the relationship between estimated ideal points and involvement in the parapolitics scandal by fitting logistic regression models to posterior samples of ideal points from both models. In both cases, we find that more extreme and coalition-aligned ideal points are significantly associated with parapolitics involvement.

This study contributes to political methodology by demonstrating the importance of geometric assumptions in latent trait modeling, showing that circular spatial voting models offer greater flexibility and robustness in fragmented or non-ideological political systems. It also provides novel insights into Colombian legislative dynamics during a period of exceptional institutional stress, offering empirical evidence that patterns of scandal involvement were not ideologically neutral but systematically aligned with the dominant coalition. 

The structure of this paper is as follows: Section~2 introduces the spatial voting models used to estimate ideal points. Section~3 describes the construction of the roll call database. Section~4 provides a descriptive analysis of the Colombian Senate during 2006–2010, with emphasis on the parapolitics scandal. Section~5 outlines the computational implementation of the models. Section~6 reports the main results, including ideal point estimates, comparisons between models, and associations with parapolitics involvement. Finally, Section~7 presents the main conclusions and outlines avenues for future research.

\section{Bayesian estimation of ideal points}

This section introduces two spatial voting models used to estimate legislators’ ideal points from roll call behavior: the Euclidean model and its circular extension. The Euclidean model assumes preferences lie in a low-dimensional linear space, with voting driven by proximity to ideological alternatives. While widely applied, this approach can misrepresent cases where ideological extremes behave similarly. To address this, we also consider a circular model that embeds ideal points on the unit circle, offering a more flexible geometric framework. We present the formulation and inference procedures for both models, emphasizing their assumptions and key differences.

\subsection {Euclidean model}

We consider \(n\) actors voting on \(m\) motions, where each vote represents a choice between two alternatives: Aye and Nay. 
The vote of the \(i\)-th actor on the \(j\)-th motion is denoted by \(y_{i,j}\), with \(i = 1, \ldots, n\) and \(j = 1, \ldots, m\), where \(1\) indicates a vote in favor and \(0\) indicates a vote against.
The voting alternatives, \( \mathbf{p}_j \) (in favor) and \( \mathbf{q}_j \) (against), for the \(j\)-th motion are represented as points in \( \mathbb{R}^d \), with the goal of obtaining a low-dimensional geometric representation often referred to as a \textit{spectrum} or \textit{political space} \citep[see][]{jackman2001multidimensional, luque2023bayesian, lipman2023explaining}.

We assume that actors are more likely to choose the option closest to their political preference (ideal point) than the option farther from it \citep{carroll2013structure}. This behavior is modeled using stochastic utility functions, which depend on both the proximity of the ideal point to the voting option and a random error term \citep{poole2005spatial}. In this way, the voting behavior of the actors is assumed to be based on their ideal points according to utility functions given by:  
\[
U_i(\mathbf{p}_j) = -\|\mathbf{p}_j - \boldsymbol{\beta}_i\|^2 + \eta_{i,j}, \qquad \text{and} \qquad 
U_i(\mathbf{q}_j) = -\|\mathbf{q}_j - \boldsymbol{\beta}_i\|^2 + \nu_{i,j}.
\]
The utilities of the \( i \)-th actor when voting in favor or against are represented by \( U_i(\mathbf{p}_j) \) and \( U_i(\mathbf{q}_j) \), respectively. It is assumed that \( \eta_{i,j} \) and \( \nu_{i,j} \) are independent random variables, such that the expected value and variance of \( \eta_{i,j} - \nu_{i,j} \) are \( 0 \) and \( \sigma_j^2 \), respectively.
The form of the utility function implicitly assumes several conditions: The only relevant factor in the political space is the distance between the ideal points and the voting options; the utility of a vote in favor (against) decreases as \( \|\mathbf{p}_j - \boldsymbol{\beta}_i\| \) (\( \|\mathbf{q}_j - \boldsymbol{\beta}_i\| \)) increases; and preferences are symmetric \citep{curini2020sage}.

The \( i \)-th actor votes in favor of the \( j \)-th roll call if and only if \( U_i(\mathbf{p}_j) > U_i(\mathbf{q}_j) \). From this, we have $\textsf{Pr}(y_{i,j} = 1 \mid \mathbf{p}_j, \mathbf{q}_j, \sigma_j, \boldsymbol{\beta}_i) = \textsf{Pr}(U_i(\mathbf{p}_j) - U_i(\mathbf{q}_j) > 0)$, and by resolving the difference \( U_i(\mathbf{p}_j) - U_i(\mathbf{q}_j) \), we obtain:  
\[
\textsf{Pr}(y_{i,j} = 1 \mid \mathbf{p}_j, \mathbf{q}_j, \sigma_j, \boldsymbol{\beta}_i) = \textsf{Pr}(\epsilon_{i,j} < \mu_j + \boldsymbol{\alpha}_j^\textsf{T} \boldsymbol{\beta}_i) = \textsf{G}(\mu_j + \boldsymbol{\alpha}_j^\textsf{T} \boldsymbol{\beta}_i),
\]
where \( \epsilon_{i,j} = (\nu_{i,j} - \eta_{i,j}) / \sigma_j \), \( \mu_j = (\mathbf{q}_j^\textsf{T} \mathbf{q}_j - \mathbf{p}_j^\textsf{T} \mathbf{p}_j) / \sigma_j \), and \( \boldsymbol{\alpha}_j = 2(\mathbf{p}_j - \mathbf{q}_j) / \sigma_j \). The parameter \( \mu_j \) represents the approval parameter, \( \boldsymbol{\alpha}_j \) is the discrimination parameter, and \( \textsf{G} \) is an appropriate link function. For mathematical details, see \cite{luque2022bayesian}.

The distribution of the random variable \( \epsilon_{i,j} \) determines the structure of the model. For instance, if \( \epsilon_{i,j} \) follows a standard normal distribution, then \textsf{G} corresponds to the cumulative distribution function of the standard normal, resulting in a probit model. In this case, we have that $y^*_{i,j} = \mu_j + \boldsymbol{\alpha}_j^\textsf{T} \boldsymbol{\beta}_i + \epsilon_{i,j}$, with $\epsilon_{i,j} \overset{\mathrm{iid}}{\sim} \textsf{N}(0,1)$, where \( y^*_{i,j} \) is the linear predictor driving the probability of a favorable vote. Consequently, we have:
\[
y_{i,j} \mid \mu_j, \boldsymbol{\alpha}_j, \boldsymbol{\beta}_i \overset{\mathrm{iid}}{\sim} \textsf{Bernoulli}(\Phi(\mu_j + \boldsymbol{\alpha}_j^\textsf{T} \boldsymbol{\beta}_i)),
\]
where $\Phi(\cdot)$ is the cumulative distribution function associated with a standard normal random variable. Note that, unlike in standard regression models, both \( \boldsymbol{\alpha}_j \) and \( \boldsymbol{\beta}_i \) are unknown parameters. The vectors \( \boldsymbol{\beta}_i \) are indexed by \( i \) because they represent characteristics of the actors, while the vectors \( \boldsymbol{\alpha}_j \) are indexed by \( j \) as they correspond to parameters specific to the roll call items.

Let \( \mathbf{Y} \) denote the \( n \times m \) matrix of observed votes, where the \( (i,j) \)-th element is \( y_{i,j} \), \( \boldsymbol{\mu} = (\mu_1, \ldots, \mu_m) \), \( \mathbf{A} \) is the \( m \times d \) matrix whose \( j \)-th row is \( \boldsymbol{\alpha}_j^\textsf{T} \), and \( \mathbf{B} \) is the \( n \times d \) matrix whose \( i \)-th row is \( \boldsymbol{\beta}_i^\textsf{T} \). Under the assumption that the \( y_{i,j} \) are exchangeable \citep{bernardo2009bayesian} given \( \mu_j \), \( \boldsymbol{\alpha}_j \), and \( \boldsymbol{\beta}_i \), we have:
\[
\textsf{p}(\mathbf{Y} \mid \boldsymbol{\mu}, \mathbf{A}, \mathbf{B}) = \prod_{i=1}^n \prod_{j=1}^m \textsf{G}(\mu_j + \boldsymbol{\alpha}_j^\textsf{T} \boldsymbol{\beta}_i)^{y_{i,j}} \left[1 - \textsf{G}(\mu_j + \boldsymbol{\alpha}_j^\textsf{T} \boldsymbol{\beta}_i)\right]^{1-y_{i,j}}.
\]

To complete the specification of the model, normal prior distributions are used for the parameters \( \mu_j \), \( \boldsymbol{\alpha}_j \), and \( \boldsymbol{\beta}_i \). Specifically, we let
\[
\mu_j, \boldsymbol{\alpha}_j \mid \boldsymbol{\alpha}_0, \mathbf{A}_0 \overset{\mathrm{ind}}{\sim} \textsf{N}(\boldsymbol{\alpha}_0, \mathbf{A}_0) \qquad \text{and} \qquad \boldsymbol{\beta}_i \mid \mathbf{b}_i, \mathbf{B}_i \overset{\mathrm{ind}}{\sim} \textsf{N}(\mathbf{b}_i, \mathbf{B}_i),
\]
where \( \boldsymbol{\alpha}_0 \), \( \mathbf{A}_0 \), \( \mathbf{b}_i \), and \( \mathbf{B}_i \) are the hyperparameters of the model. The vectors \( \boldsymbol{\alpha}_0 \) and \( \mathbf{b}_i \) represent prior means, while \( \mathbf{A}_0 \) and \( \mathbf{B}_i \) are their corresponding covariance matrices. Common choices for these hyperparameters are \( \boldsymbol{\alpha}_0 = \mathbf{0} \), \( \mathbf{A}_0 = 25\mathbf{I} \), \( \mathbf{b}_0 = \mathbf{0} \), and \( \mathbf{B}_i = \mathbf{I} \) \citep{clinton2004statistical}. 

The primary goal is to estimate the ideal points \( \boldsymbol{\beta}_1, \ldots, \boldsymbol{\beta}_n \) to uncover and describe the latent traits underlying voting behavior. Details of the parameter estimation process can be found in \cite{luque2022bayesian}, where Markov chain-based methods are presented to explore the posterior distribution given by \( \textsf{p}(\mathbf{Y} \mid \boldsymbol{\alpha}, \mathbf{A}, \mathbf{B}) \propto \textsf{p}(\mathbf{Y} \mid \boldsymbol{\alpha}, \mathbf{A}, \mathbf{B}) \, \textsf{p}(\boldsymbol{\alpha}, \mathbf{A}, \mathbf{B}) \).

\subsection {Circular model}

Traditional spatial voting models typically embed latent preferences in low-dimensional Euclidean space to estimate ideal points and interpret legislative behavior. This geometric assumption has proven useful for uncovering ideological patterns but may fail in contexts where ideological extremes exhibit similar voting behavior for fundamentally different reasons \citep{spirling2007uk, spirling2010identifying}. Such behavior results in distorted spatial representations when using Euclidean distance, often misclassifying ideological extremists as centrists. To address this limitation, \citet{yu2020spherical, yu2021spatial} introduced the \textit{Spherical Latent Factor Model}, a flexible generalization of spatial voting models in which latent preferences are embedded in the unit circle or hypersphere, rather than flat Euclidean space.

In the one-dimensional spherical model, the ideal points \( \beta_i \) and the bill parameters \( \psi_j \) and \( \zeta_j \), corresponding to the ``Yea'' and ``Nay'' positions respectively, are represented as angular coordinates on the circle, taking values in the interval \( [-\pi, \pi] \). The model replaces Euclidean distances with geodesic distances on the unit circle, leading to the following utility functions for actor \( i \) voting on motion \( j \):
\[
U_i(\psi_j)  = -(\arccos(\cos(\psi_j - \beta_i)))^2 + \epsilon_{i,j} \qquad \text{and} \qquad 
U_i(\zeta_j) = -(\arccos(\cos(\zeta_j - \beta_i)))^2 + \nu_{i,j},
\]
where \( \epsilon_{i,j} \) and \( \nu_{i,j} \) are random shocks. The actor votes in favor if and only if \( U_i(\psi_j) > U_i(\zeta_j) \), which leads to $\textsf{Pr}(y_{i,j} = 1 \mid \beta_i, \psi_j, \zeta_j, \kappa_j) = \textsf{G}_{\kappa_j}(z_{i,j})$, with
\[
z_{i,j} = (\arccos(\cos(\zeta_j - \beta_i)))^2 - (\arccos(\cos(\psi_j - \beta_i)))^2,
\]
and \( \textsf{G}_{\kappa_j} \) denoting the cumulative distribution function of a scaled, symmetric Beta distribution on \( [-\pi^2, \pi^2] \):
\[
\textsf{g}_{\kappa_j}(z) = \frac{1}{2\pi^2} \cdot \frac{\Gamma(2\kappa_j)}{\Gamma(\kappa_j)^2} \left( \frac{\pi^2 + z}{2\pi^2} \right)^{\kappa_j - 1} \left( \frac{\pi^2 - z}{2\pi^2} \right)^{\kappa_j - 1}.
\]

To perform Bayesian inference, prior distributions are assigned to all parameters \citep{yu2021spatial}. The latent angular positions \( \beta_i, \psi_j, \zeta_j \) follow von Mises distributions:
\[
\beta_i \mid \omega_\beta \sim \textsf{vonMises}(0, \omega_\beta), \qquad
\psi_j \sim \textsf{vonMises}(0, \omega_\psi), \qquad
\zeta_j \sim \textsf{vonMises}(0, \omega_\zeta),
\]
where the von Mises density is
\[
\textsf{p}(z \mid \mu, \omega) = \frac{1}{2\pi I_0(\omega)} \exp\{\omega \cos(z - \mu)\}, \qquad z \in [-\pi, \pi],
\]
and \( I_0(\omega) \) is the modified Bessel function of order zero. The concentration parameter \( \omega \) controls the dispersion (the von Mises distribution approximates a normal distribution for large \( \omega \)). A Gamma prior is placed on \( \omega_\beta \), specifically \( \omega_\beta \sim \textsf{Gamma}(1, 1/10) \), while \( \omega_\psi \) and \( \omega_\zeta \) are fixed at zero to yield noninformative priors. The concentration parameters of the link function, \( \kappa_j \), are modeled hierarchically as \( \kappa_j \mid \beta_\kappa \overset{\text{ind}}{\sim} \textsf{Gamma}(1, \beta_\kappa) \), with a hyperprior \( \beta_\kappa \sim \textsf{Gamma}(1, 25) \).

This spherical formulation addresses key limitations of Euclidean models by eliminating scale-related identifiability issues through bounded angular distances. Moreover, as noted by \citet{tversky196919intransitivity} and \citet{weisberg1974dimensionland}, spherical geometry naturally captures horseshoe-shaped ideological structures, where political extremes lie close in curved space rather than at opposite ends of a line. Importantly, the spherical model recovers the Euclidean model as a limiting case: When angular parameters are tightly concentrated around zero and the link function precision \( \kappa_j \) is large, geodesic distances closely approximate their Euclidean counterparts.

For posterior inference, \citet{yu2020spherical} develop two Hamiltonian Monte Carlo (HMC) algorithms \citep[see also][]{gelman2013bayesian}. The first adapts standard HMC by incorporating von Mises-distributed momenta and modifying the leapfrog integrator to accommodate circular geometry. The second is a geodesic HMC method \citep{byrne2013geodesic}, which leverages closed-form geodesic flows on the sphere to efficiently explore curved latent spaces. These approaches enable scalable and accurate posterior sampling, even in high-dimensional non-Euclidean settings. In practice, the step size as well as the number of leapfrog steps are tuned to achieve acceptance rates between 60\% and 90\%.

\section{Data}

Recent research highlights that access to roll call data in Latin America is a relatively recent development \citep{luque2024operationalizing}. In Colombia, systematic collection of such data began in 2006 \citep{carroll2016unrealized}. Analyzing these records offers valuable and novel insights into legislative behavior and broader political dynamics in the region \citep{mcdonnell2017formal, luque2023bayesian}.

We obtained roll call data from \textit{Congreso Visible} (\url{https://congresovisible.uniandes.edu.co/}) and subsequently normalized and structured it following a systematic methodology inspired by \citet{luque2023bayesian}. Compiling and organizing roll call data is a complex process that involves manually extracting and reconciling information from multiple external sources \citep{morgenstern2003patterns, luque2023bayesian, luque2024operationalizing}. In particular, we consulted \texttt{Wikipedia} to identify active senators, their political affiliations, party switches, resignations, and involvement in the parapolitics scandal. While \texttt{Wikipedia} is not an official source, it remains one of the most comprehensive and accessible repositories of information on the Colombian Congress, especially for non-current legislative periods. Notably, the official website of the Colombian Congress does not maintain historical archives of past legislatures.

To ensure the accuracy of the collected data, we cross-referenced the information with multiple sources, including records from \textit{Congreso Visible} and reports on legislators published in Colombian print and digital media such as \textit{Revista Semana}, \textit{El Tiempo}, and \textit{El Espectador}. As a result, we consolidated a comprehensive database comprising 147 senators, including both principal members and substitutes who assumed office due to death, dismissal, or other circumstances, and 136 roll call votes from plenary sessions of the Colombian Senate during the 2006–2010 legislative period.

\section{Colombian Senate 2006-2010 and parapolitics}

The 2006--2010 Senate operated during the second presidential term of Álvaro Uribe Vélez, a period marked by a pivotal political juncture in Colombia: The emergence of the \textit{parapolitics scandal} \citep{bakiner2020endogenous, pardo2020dual}. This scandal revealed connections between elected officials and illegal far-right paramilitary groups, organizations involved in drug trafficking and clandestinely linked to public institutions amid the Colombian armed conflict \citep{avila2012politics}. As a consequence, the Senate underwent significant changes in its composition due to numerous resignations and dismissals of legislators implicated in the scandal \citep{staff2007more}.

In the political landscape of the 2006–2010 Senate, most parties supported President Álvaro Uribe’s democratic security agenda. The governing coalition was composed of \textit{ Partido de la U} (PU), \textit{Partido Conervador Colombiano} (PC), \textit{Cambio Radical} (CR), \textit{Colombia Viva} (CV), \textit{Convergencia Ciudadana} (CC), \textit{Alas Equipo Colombia} (AEC), and \textit{Colombia Democrática} (CD). In opposition were the \textit{Polo Democrático Alternatico} (PDA) and the \textit{Partido Liberal Colombiano} (PL), while the \textit{Movimiento MIRA} maintained an independent position. Minority representation in the Senate included the \textit{Autoridades Indígenas de Colombia} (AICO) and the \textit{Alianza Social Indígena} (ASI). The parties with the lowest proportions of senators implicated in the parapolitics scandal were the PC (25.0\%), PL (27.3\%), and PU (37.9\%). In contrast, the highest rates of implication were observed among senators from CV (66.7\%), AEC (50.0\%), and CD (50.0\%). The remaining parties had no members linked to the scandal.

\begin{table}[!htb]
\centering
\begin{tabular}{lcc}
\hline
Party & Number of seats & Number of senators  \\ \hline
\textit{Alas Equipo Colombia} & 5 & 8  \\ 
\textit{Movimiento MIRA} & 2 & 2  \\ 
\textit{Partido Conservador Colombiano} & 18 & 24  \\ 
\textit{Convergencia Ciudadana} & 7 & 11  \\ 
\textit{Partido de la U} & 20 & 29  \\ 
\textit{Movimiento Colombia Viva} & 2 & 6  \\ 
\textit{Polo Democrático Alternativo} & 10 & 12  \\ 
\textit{Colombia Democrática} & 3 & 8  \\ 
\textit{Partido Liberal Colombiano} & 18 & 22  \\ 
\textit{Cambio Radical} & 15 & 23  \\ 
\textit{Alianza Social Indígena} & 1 & 1  \\ 
\textit{Autoridades Indígenas de Colombia} & 1 & 1  \\ \hline
\end{tabular}
\caption{Number of seats and number of senators by party.}
\label{tab:table0}
\end{table}

\begin{figure}[!b]
    \centering
    \subfigure[Raw data.] {\includegraphics[scale=0.44]{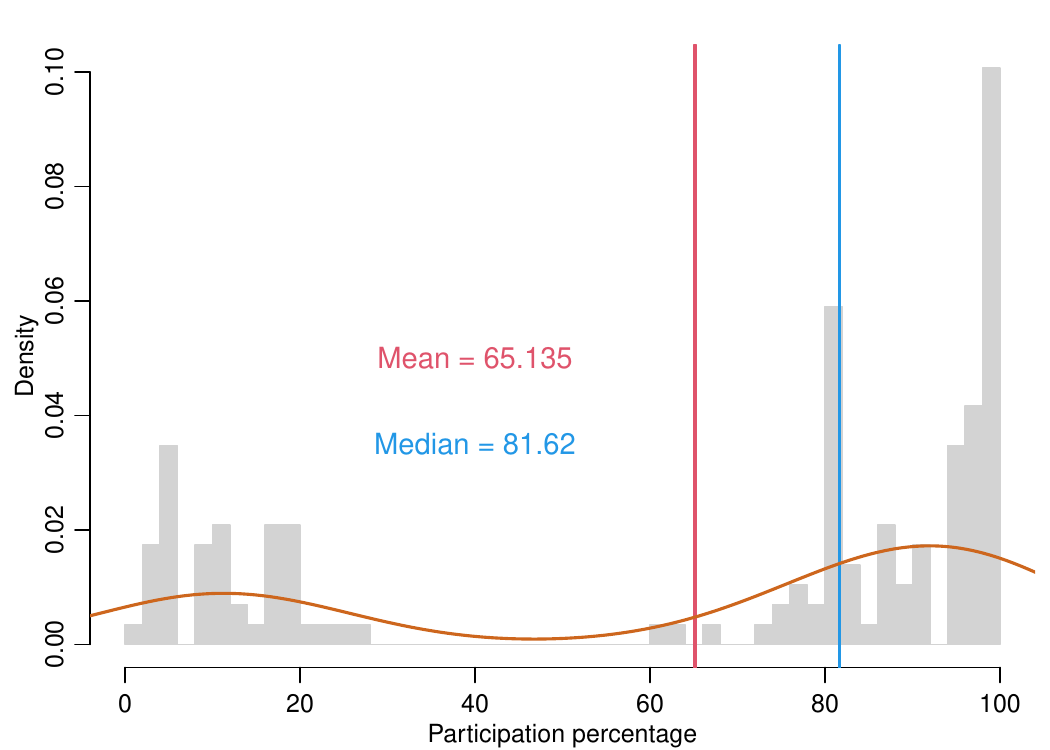}}
    \subfigure[By party.] {\includegraphics[scale=0.44]{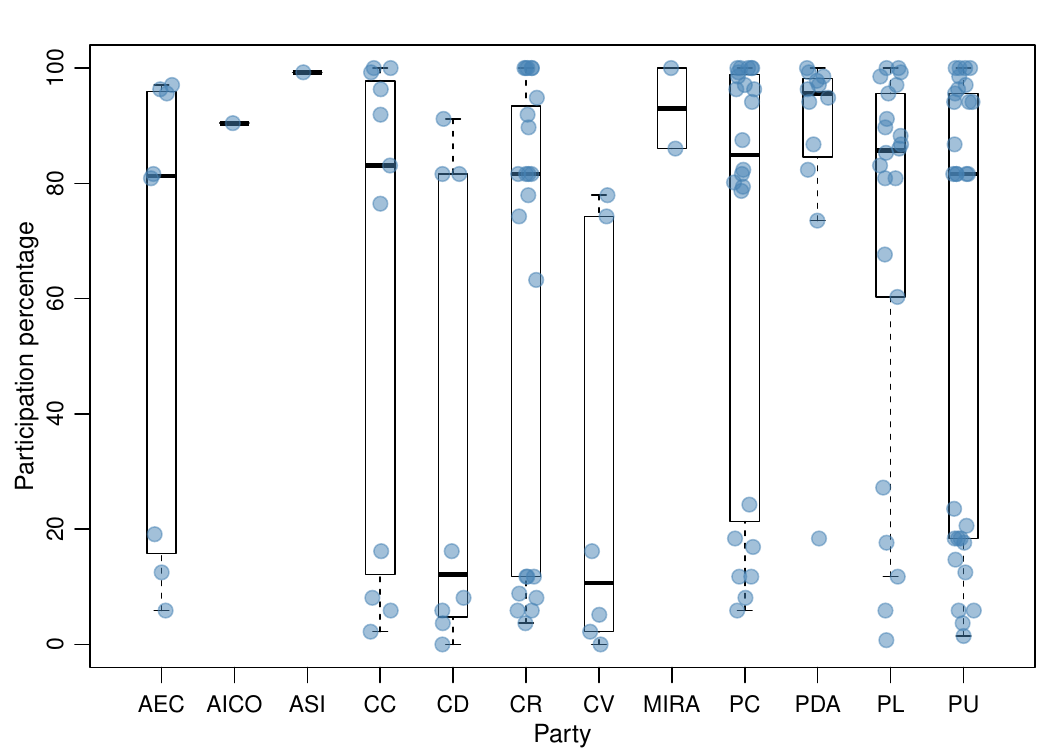}}
    \caption{Percentage of participation.}
    \label{fig_histpart_boxpart}
\end{figure}

During this period, the dynamics of the legislative body reflect significant changes in its composition and patterns of participation. Table \ref{tab:table0} reports the distribution of Senate seats by political party, detailing both the number of seats and the total number of senators who held office between 2006 and 2010. Discrepancies between these figures arise from replacements within the chamber, with some seats occupied by up to three different senators over the course of the term. In terms of legislative activity, Figure \ref{fig_histpart_boxpart} displays senators’ voting participation rates, revealing considerable heterogeneity, from very low engagement to consistently high involvement in roll call votes. Low participation is largely attributable to resignations, substitutions, temporary replacements, or deaths. Additionally, Figure \ref{fig_histpart_boxpart} shows that voting participation varies systematically across political parties.

\begin{figure}[!h]
    \centering
    \subfigure[Raw data.] {\includegraphics[scale=0.44]{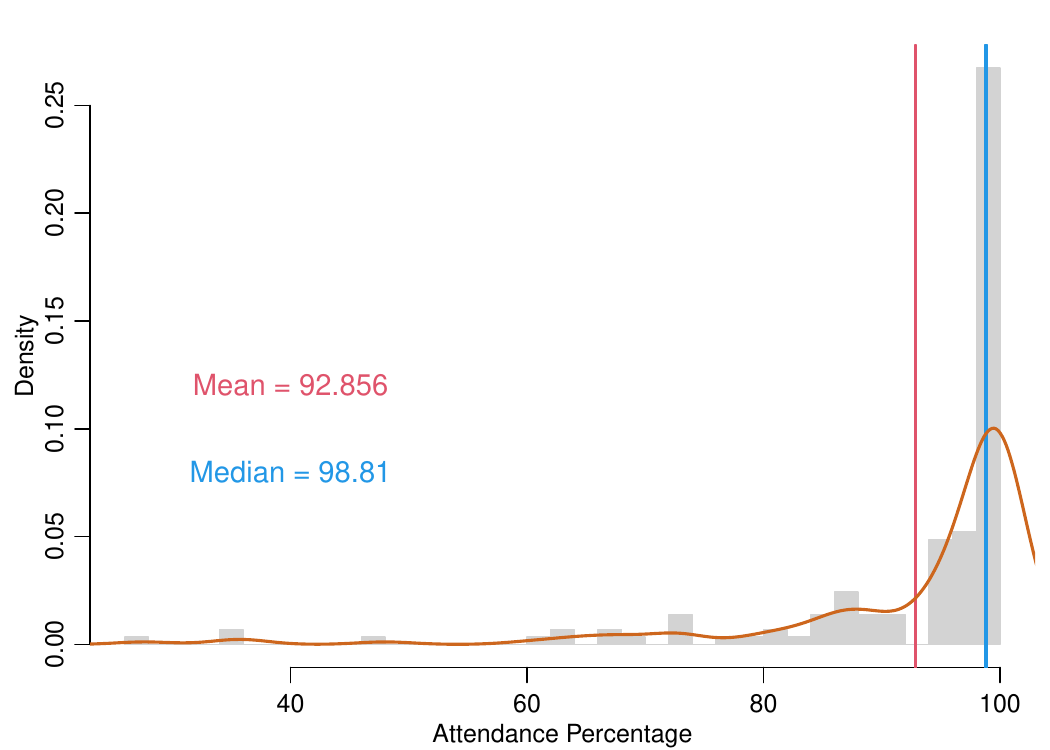}}
    \subfigure[By party.] {\includegraphics[scale=0.44]{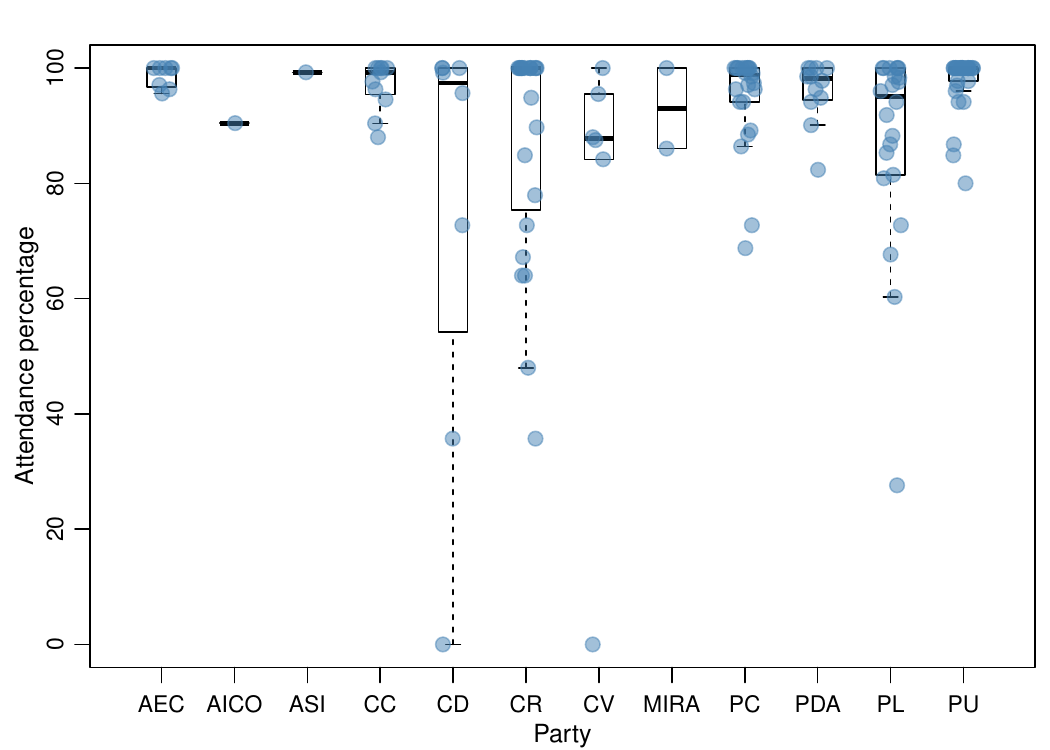}}
    \caption{Percentage of attendance.}
    \label{fig_histasis_boxasist}
\end{figure}

\begin{figure}[!b]
    \centering
    \subfigure[Attendance.] {\includegraphics[scale=0.44]{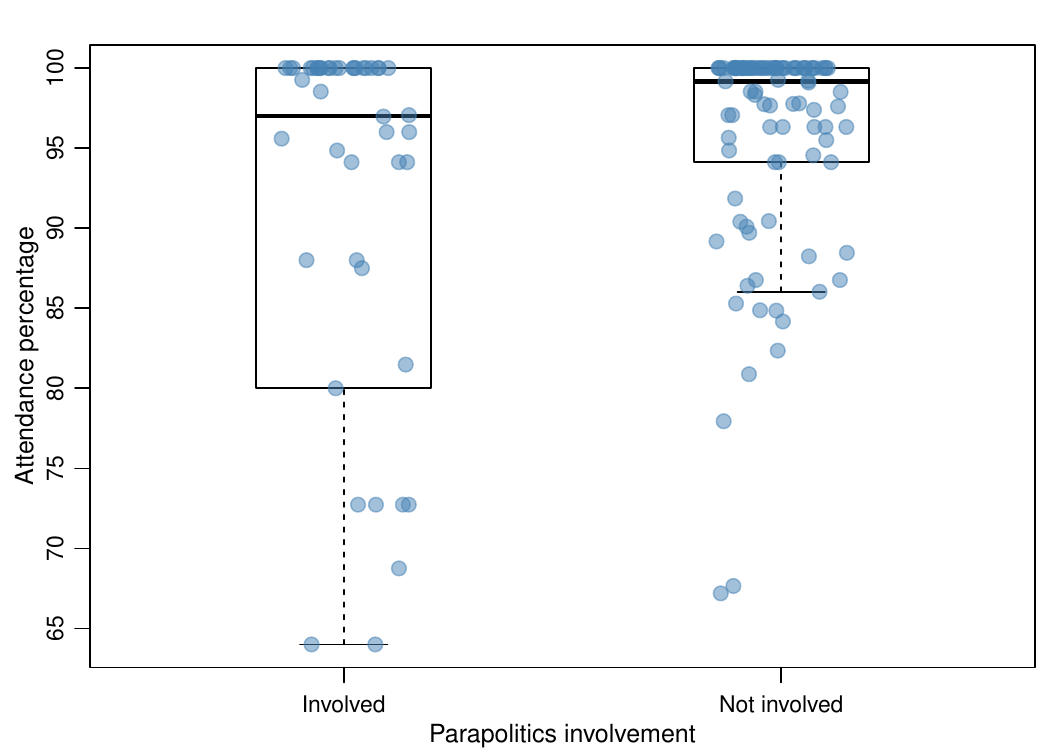}}
    \subfigure[Abstention.] {\includegraphics[scale=0.44]{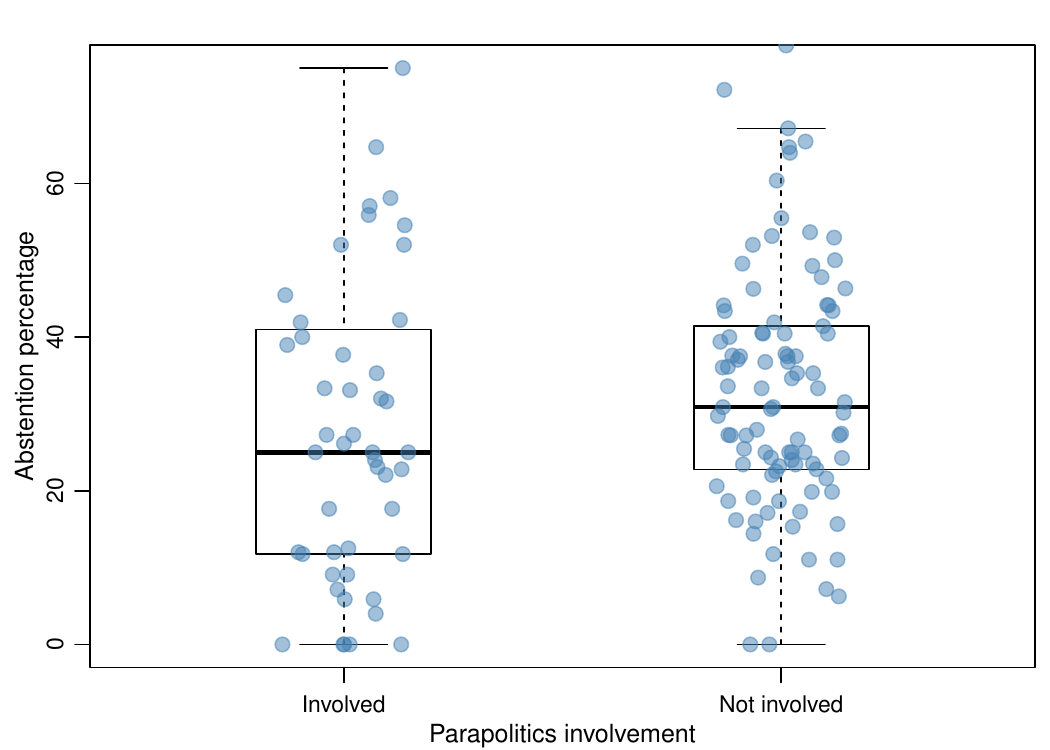}}
    \caption{Attendance and abstention rates by parapolitics involvement.}
    \label{fig_boxasispar_boxabstpar}
\end{figure}

Figure~\ref{fig_histasis_boxasist} presents senators’ attendance rates, showing that over 50\% of legislators attended more than 98\% of roll call votes. At the party level, most groups exhibit stable attendance rates above 80\%; however, greater variability is observed within CD and CR. Additionally, Figure~\ref{fig_boxasispar_boxabstpar} explores the relationship between involvement in the parapolitics scandal and legislative attendance. Attendance rates among implicated senators are more widely dispersed and generally lower compared to their non-implicated counterparts. Both distributions exhibit negative skewness, aligning with the overall pattern of high attendance concentrated near the upper end of the scale.

\begin{figure}[!htb]
    \centering
    \subfigure[Raw data.] {\includegraphics[scale=0.44]{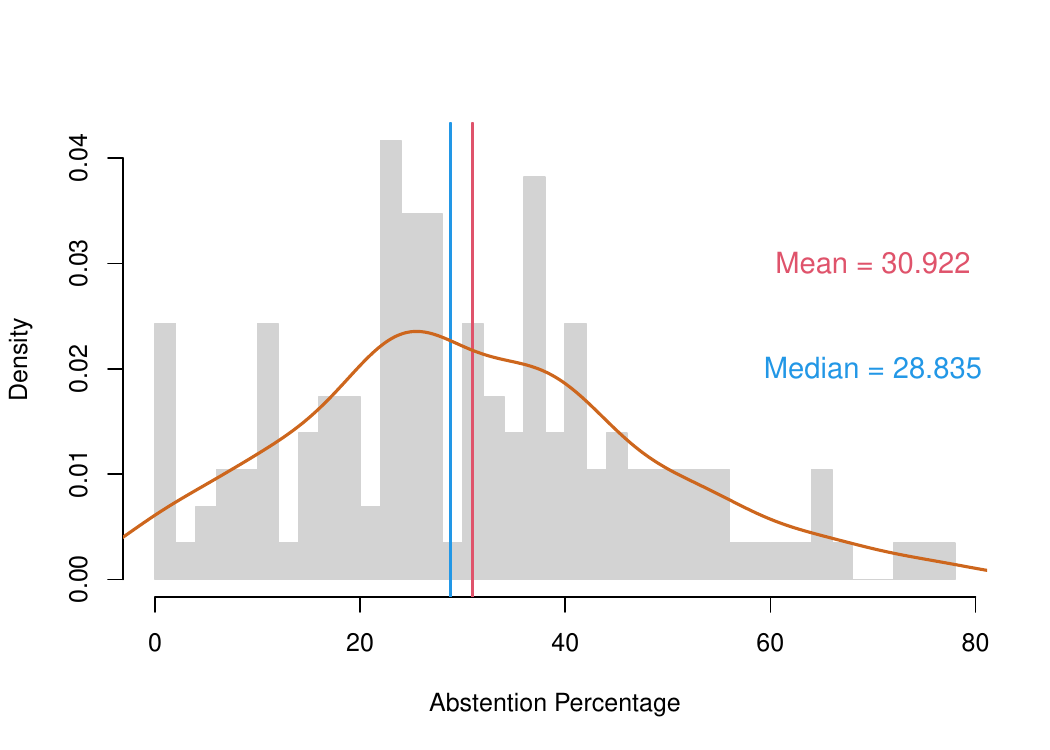}}
    \subfigure[By party.] {\includegraphics[scale=0.44]{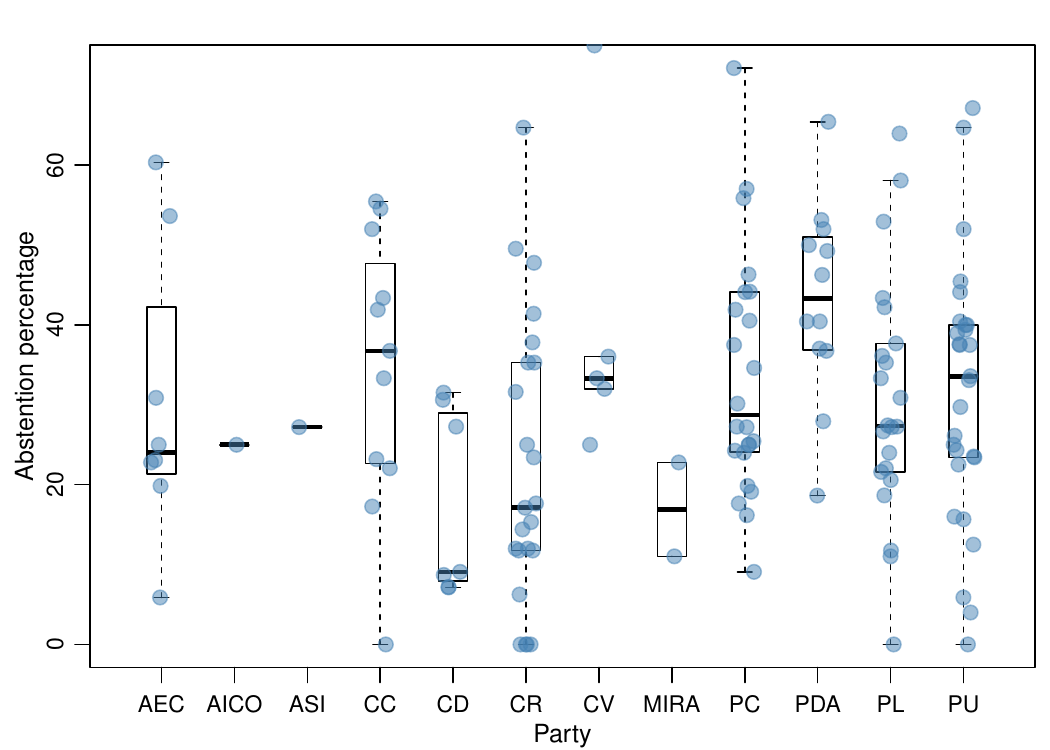}}
    \caption{Percentage of abstention.}
    \label{fig_histabst_boxabst}
\end{figure}

Finally, patterns of voting abstention also reveal notable distinctions. Figure~\ref{fig_histabst_boxabst} shows that approximately 72\% of senators maintain abstention rates at or below 40\%, while only 7\% reach levels of 60\% or higher. At the party level, Figure~\ref{fig_histabst_boxabst} indicates that most parties exhibit average abstention rates below 60\%, with MIRA standing out for its minimal variability. In contrast, the main opposition party, PDA, records the highest levels of abstention. Figure~\ref{fig_boxasispar_boxabstpar} compares abstention rates between senators implicated and not implicated in the parapolitics scandal, revealing no substantial differences between the two groups. In both cases, abstention percentages range from 0\% to 40\%, though the non-implicated group exhibits a slightly higher median.

\section{Computation}

This section presents the computational implementation of the two latent factor models used to estimate legislators' ideal points. First, we describe the Euclidean model, which assumes a one-dimensional ideological space and relies on standard spatial modeling techniques. We then detail the spherical model, which also operates in one dimension but embeds the ideal points on the unit circle to account for the geometric limitations of Euclidean space. For each model, we outline the prior specifications, MCMC settings, and procedures used to ensure identifiability. We also discuss the projection of spherical estimates onto a linear space to enable direct comparison between the two approaches. All code used to reproduce our results is available at \url{https://github.com/jstats1702/ideal-parapolitics}.

\subsection{Euclidean model}

For the Euclidean model, we adopt a one-dimensional specification to estimate legislators' political preferences, setting \( d = 1 \). In this framework, the parameters \(\boldsymbol{\alpha}_j\) (for \( j = 1, \ldots, m \)) and \(\boldsymbol{\beta}_i\) (for \( i = 1, \ldots, n \)) are scalars, where \( m = 136 \) corresponds to the number of plenary voting lists and \( n = 144 \) to the number of senators. Three senators, Jorge de Jesús Castro Pacheco (CV), Álvaro Alfonso García Romero (CD), and Jairo Enrique Merlano Fernández (PU), are excluded from the analysis due to the absence of plenary voting records.

To ensure model identifiability, we fix the positions of two senators (commonly referred to as anchors) at the endpoints of the ideological spectrum, specifically at 1 and -1. For this purpose, we select Guillermo Alfonso Jaramillo Martínez (PDA) and Carlos Cárdenas Ortiz (PU) as representatives of ideologically opposing parties \citep{clinton2004statistical, luque2023bayesian}. Additionally, following the approach of \citet{luque2023bayesian} and others, we set the hyperparameter values to \( \alpha_0 = 0 \) and \( A_0 = 25 \), and \( b_i = 0 \) and \( B_i = 1 \). This choice yields uninformative priors for the parameters \( \mu_j \) and \( \boldsymbol{\alpha}_j \) while ensuring that, a priori, over 99\% of the ideal points fall within the interval \([-3,3]\) \citep{luque2021metodos}.

We use the logit function as the link, noting that the probit link produces similar results in univariate settings \citep{luque2023bayesian}. To estimate the model, we implement a Markov chain Monte Carlo (MCMC) algorithm \citep{gamerman2006markov, gelman2013bayesian} to sample from the posterior distribution. As initial values for model fitting, we assign a value of 1 to the ideal points of coalition senators and -1 to those of opposition senators.

The model is implemented in \texttt{R}  using \texttt{STAN} (version 2.26.1). A total of 80,000 iterations are performed, with the first 16,000 used as warm-up. To reduce the influence of initial values and minimize autocorrelation, we apply systematic thinning, retaining every fifth iteration, yielding 12,800 post-warm-up samples (64,000/5) for inference. Model convergence is assessed, and no lack of convergence is detected in any scenario.

\subsection{Circular model}

For the circular model, we also adopt a one-dimensional specification, where legislators' ideal points \( \beta_i \) are constrained to lie on the unit circle, taking values within the interval \( [-\pi, \pi] \). The parameters \( \beta_i \), along with the bill parameters \( \psi_j \) and \( \zeta_j \), are modeled using von Mises distributions with respective concentration parameters \( \omega_\beta, \omega_\psi, \omega_\zeta \). Following \cite{yu2020spherical}, the hyperparameters are specified as follows: \( \omega_\beta \sim \textsf{Gamma}(1, 1/10) \), while \( \omega_\psi \) and \( \omega_\zeta \) are fixed at zero to induce noninformative priors for the bill parameters. The concentration parameters of the link function, \( \kappa_j \), are modeled hierarchically as \( \kappa_j \mid \beta_\kappa \overset{\text{ind}}{\sim} \textsf{Gamma}(1, \beta_\kappa) \), with a hyperprior \( \beta_\kappa \sim \textsf{Gamma}(1, 25) \). This setup ensures diffuse priors for discrimination parameters while allowing hierarchical regularization of the link function’s precision.

The model is implemented in \texttt{R} using the \texttt{SLFM1D} package \citep{yu2021spatial}. Model fitting is carried out via Hamiltonian Monte Carlo methods specifically adapted to the geometry of the unit circle. In particular, we employ both von Mises momentum-based HMC and geodesic HMC, which exploit the manifold structure of the circular space to efficiently explore the posterior distribution. A total of 30,000 MCMC iterations are performed, with the first 10,000 discarded as burn-in. Convergence diagnostics indicate no evidence of lack of convergence. Since the ideal points \( \beta_i \) lie on the circle, the model is inherently non-identifiable up to rotation, reflection, and translation. To resolve these ambiguities, we apply a post-processing alignment procedure to the posterior samples.

To resolve translation invariance, we align the posterior samples by fixing the location of a reference legislator at \( \pi/2 \). Specifically, Senator Carlos Cardenas Ortiz (PU) is chosen as the reference, and for each posterior draw, the angular shift required to align their estimated ideal point to \( \pi/2 \) is computed and applied to all samples. After this rotation, reflection invariance is resolved by enforcing a consistent ideological orientation. This is achieved by selecting a second reference legislator, Guillermo Alfonso Jaramillo Martinez (PDA), and ensuring that their ideal point remains positive. If, after alignment, their estimated position is negative, all samples are reflected by multiplying them by -1. This two-step procedure ensures that the estimated ideal points remain consistently oriented across all posterior draws, eliminating ambiguities in interpretation.

To enable direct comparison with the Euclidean model and assess the association between ideal points and the parapolitics scandal, we project the identified circular ideal points onto the tangent space at \( \pi/2 \) using the log map transformation. This transformation, defined as \( p_i^{(b)} = \tan(\beta_i^{(b)} - \pi/2) \), maps the circular latent traits onto a linear space while preserving local geometric relationships. By implementing this alignment and projection strategy, we ensure that the estimated ideal points are identifiable and interpretable while preserving the geometry of the spherical model. The final aligned estimates provide a robust representation of latent political preferences, facilitating meaningful comparisons across different methodological frameworks.

Finally, as an additional check, we evaluate the impact of missing data on ideal point estimation. For this purpose, we conduct two parallel experiments. First, we fit the Euclidean model using only complete cases, that is, by removing all observations with missing votes. In contrast, for the circular model, we incorporate the missing data into the estimation process under the assumption that they are missing at random, treating them as latent variables and imputing their values through the posterior predictive distribution during model training. The results obtained from both strategies are consistent, indicating that the estimated ideal points are not substantially perturbed by the choice of missing data handling. This finding reinforces the robustness of our estimates and aligns with the results of \citet{luque2023bayesian}, who demonstrate through extensive simulation studies that ideal point estimation is stable even when missing votes are excluded. Their findings suggest that, under typical voting patterns and missingness structures, the inferred latent traits remain largely invariant regardless of whether missing data are imputed or omitted. 

\section{Results}

\begin{figure}[!htb]
    \centering
    \includegraphics{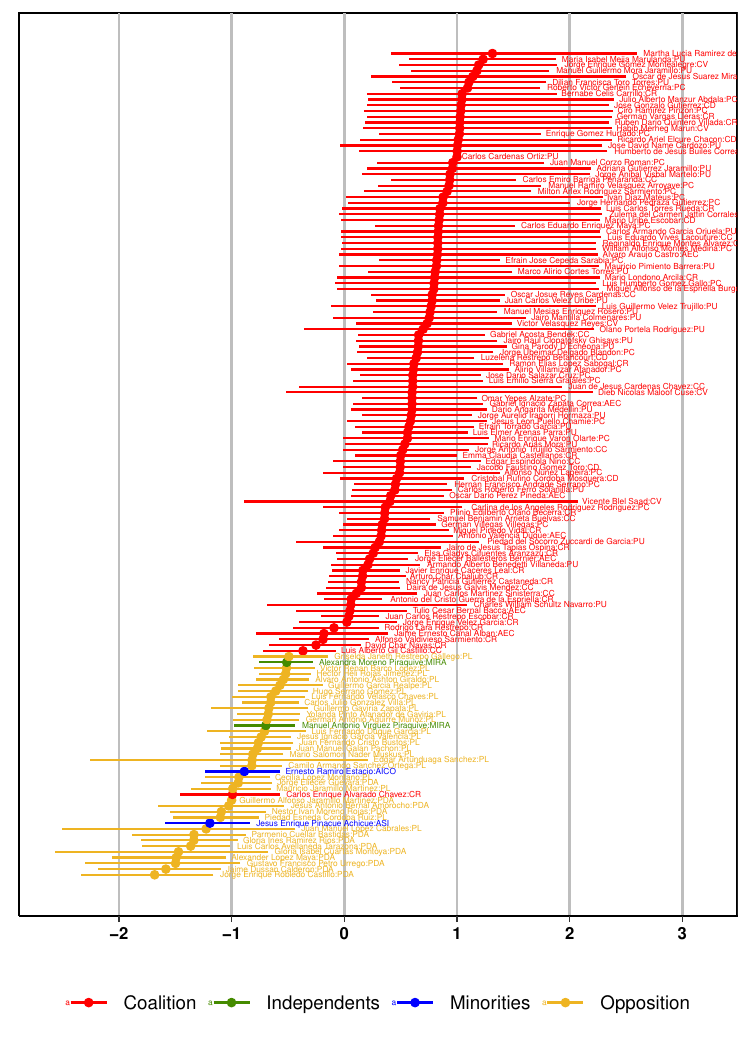}
    \caption{Estimated ideal points based on anchor legislators positioned at opposite ends of the spectrum. Points indicate posterior means, and horizontal lines represent symmetric 95\% credibility intervals computed from posterior percentiles.}
    \label{credibilidad}
\end{figure}

Figure~\ref{credibilidad} displays the estimated ideal points and corresponding credibility intervals for the 144 senators. Credibility intervals are not shown for the anchor legislators, whose positions are fixed by construction. The figure reveals a clear separation, with opposition, independent, and minority senators positioned on the left side of the political space, in contrast to most members of the governing coalition. Notably, 94.3\% of coalition senators have ideal point estimates greater than zero. The interpretation of ``left'' and ``right'' is not intrinsic but rather a consequence of the positional choices made for the anchor legislators.

The posterior mean of senators’ ideal points ranges from \(-1.68\) to \(1.31\). Table~\ref{tab:table2} reports the minimum, maximum, and posterior coefficient of variation (CV) of ideal points within each political group. These summary measures indicate that independent and minority groups display the greatest internal consistency, while the coalition and opposition exhibit higher heterogeneity. This contrasts with the 2010–2014 Senate, where the opposition and coalition were the most cohesive blocs \citep{luque2023bayesian}. Among the four groups, the governing coalition shows the highest dispersion in members’ ideological positions.

\begin{table}[!htb]
	\centering
\begin{tabular}{lcccc}
\hline
Statistic & Coalition & Independents & Minorities & Opposition \\ \hline
Minimum & -0.99 & -0.70 & -1.19 & -1.68 \\ 
Maximum & 1.31 & -0.51 & -0.89 & -0.49 \\ 
CV($\%$) & 63.95 & 21.71 & 20.80 & 37.07 \\ \hline
\end{tabular}
\caption{Summary statistics of the estimated ideal points by political group, including the minimum, maximum, and coefficient of variation (CV) for each group.}
\label{tab:table2}
\end{table}

Table~\ref{tab:table3} reports the minimum, maximum, and posterior coefficient of variation (CV) of the estimated ideal points by political party. CV values are not reported for AICO and ASI, as each has only one representative. Among the independent and opposition parties, MIRA, PDA, and PL exhibit the lowest within-party variability in ideal points. In contrast, CR, AEC, and CC stand out for their high internal heterogeneity within the governing coalition. This volatility may reflect the transformation of the Colombian party system between 2002 and 2010, shaped by the political reforms of 2003 and 2009 and the introduction of presidential reelection. These changes encouraged party switching and led some parties to incorporate politically diverse actors under a single label \citep{castro2011transformaciones}.

\begin{table}[!htb]
\centering
\scalebox{0.8}{
\begin{tabular}{lcccccccccccc}
\hline
Statistic & PDA & MIRA & AEC & CC & CD & CR & CV & PC & PU & PL & AICO & ASI \\ \hline
Minimum & -1.68 & -0.70 & -0.18 & -0.37 & 0.46 & -0.99 & 0.41 & 0.34 & 0.06 & -1.23 & -0.89 & -1.19 \\
Maximum & -0.94 & -0.51 & 1.14 & 0.94 & 1.04 & 1.05 & 1.19 & 1.09 & 1.31 & -0.49 & -0.89 & -1.19 \\ 
CV ($\%$) & 18.75 & 21.71 & 99.89 & 82.42 & 30.59 & 145.86 & 40.12 & 29.89 & 39.11 & 25.75 & NA & NA \\ \hline
\end{tabular}}
\caption{Summary measures of the estimated ideal points by political party. For each party, the minimum, maximum, and coefficient of variation (CV) are reported. The AICO and ASI parties do not have CV values as they had only one senator during this term.}
\label{tab:table3}
\end{table}

The widest credibility intervals in Figure~\ref{credibilidad} correspond to senators with very low participation rates (below 10\%), such as Vicente Blel (CV), Dieb Maloof (CV), Olano Portela (PU), Edgar Artunduaga (PL), and Luis Guillermo Vélez, whose respective participation rates were 2.21\%, 5.15\%, 3.68\%, 0.74\%, and 5.88\%. Wider credibility intervals indicate greater uncertainty in the estimated ideal points. As noted in previous studies, these intervals also tend to widen as ideal points deviate from the center of the political space \citep{luque2023bayesian}. Additionally, coalition senators generally exhibit broader intervals, reflecting greater heterogeneity and lower voting consistency within the group.

\begin{figure}[!htb]
 \centering
  \subfigure[Alexander López Maya.]{\includegraphics[width=0.4\textwidth]{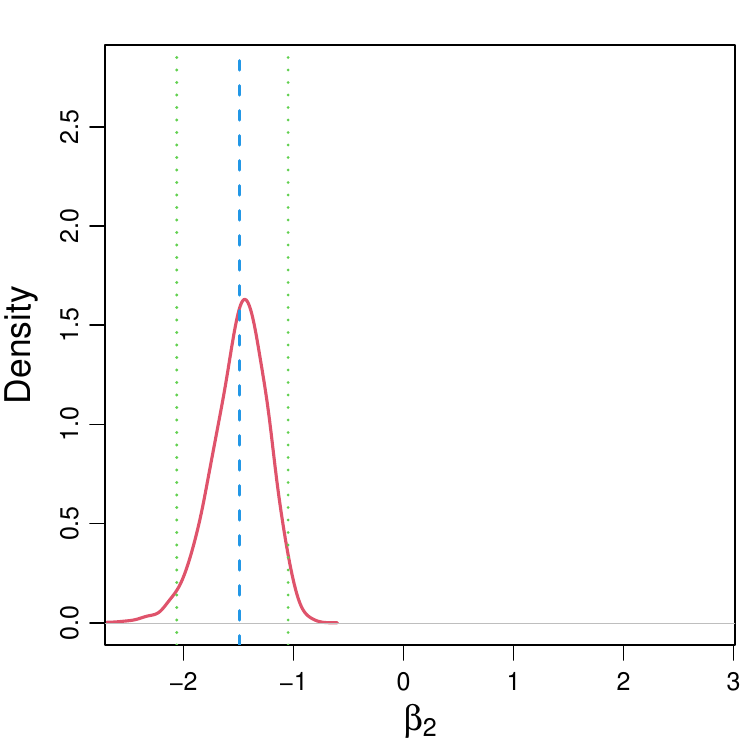}}   \label{11}
  \subfigure[Cecilia López Montaño.]{\includegraphics[width=0.4\textwidth]{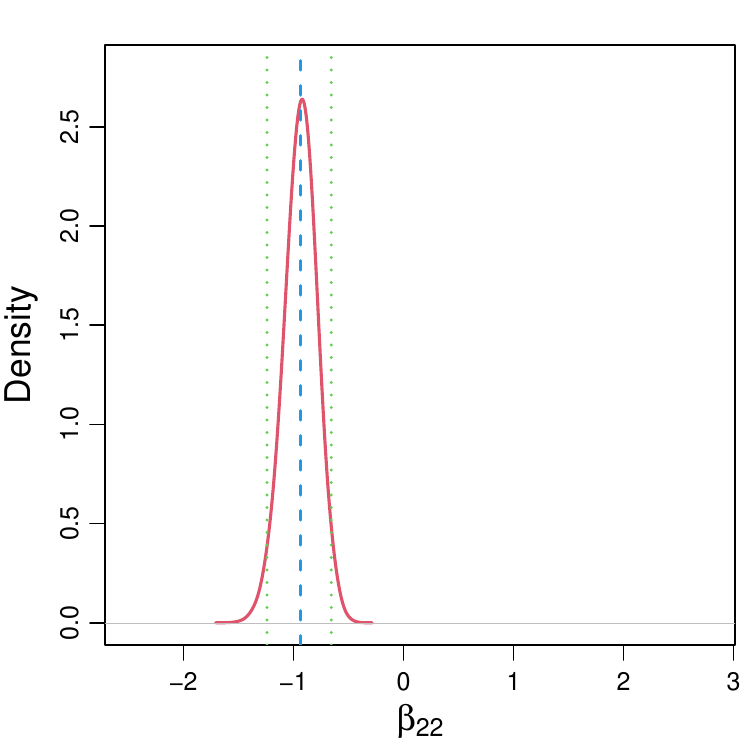}}  \label{12} \\
  \subfigure[ Mario Uribe Escobar.]{\includegraphics[width=0.4\textwidth]{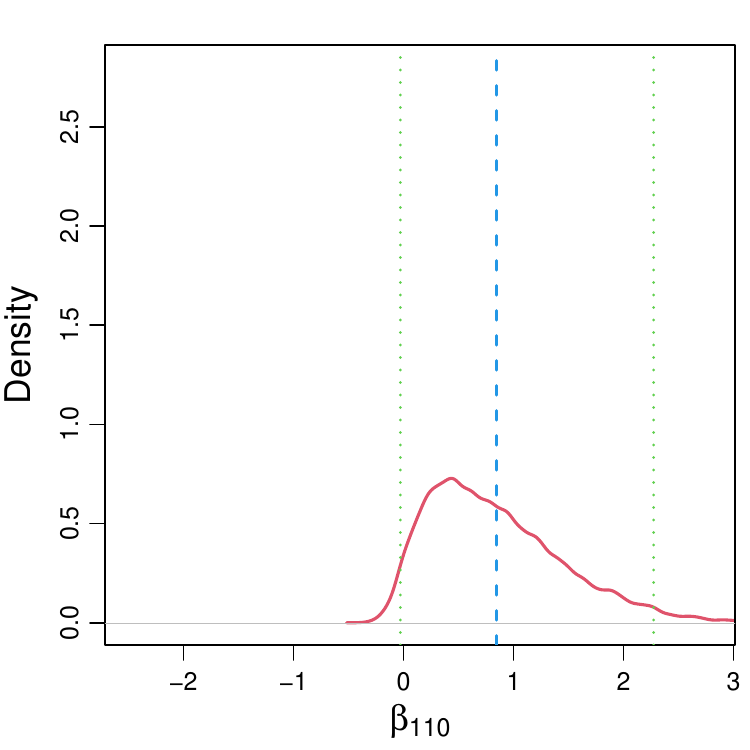}} \label{21}
  \subfigure[Zulema del Carmen Jattin Corrales.]{\includegraphics[width=0.4\textwidth]{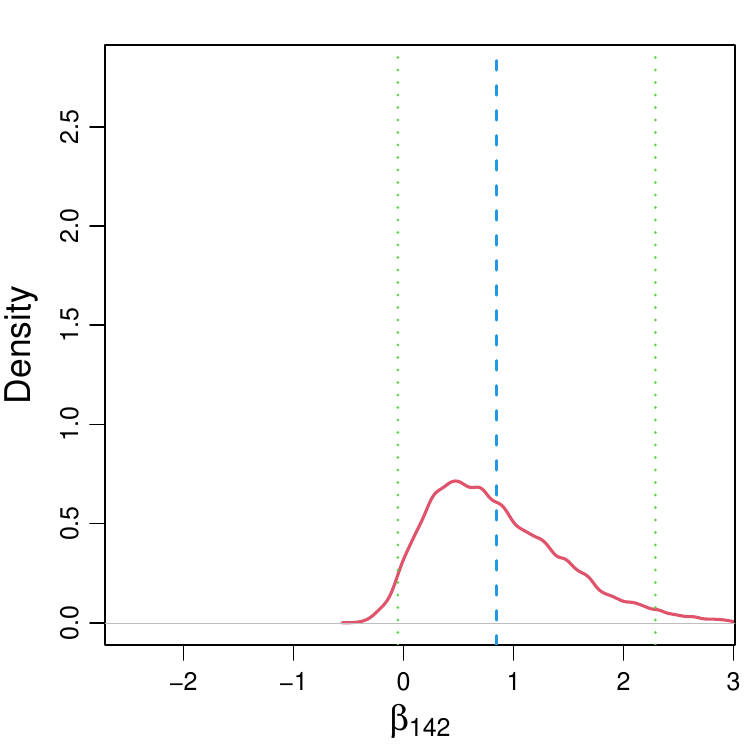}} \label{22}
 \caption{Posterior marginal distributions for four selected senators. The blue line represents the posterior mean, and the green lines indicate the 95\% credibility intervals.} 
 \label{densidades}              
\end{figure}

Of the 144 estimated ideal points, 94 (65.3\%) are significantly different from zero. Notably, 49 of the 50 ideal points indistinguishable from zero (98\%) correspond to senators from the governing coalition. Among these, 16 are members of CR, representing 32\% of the indistinguishable cases. At least half of the senators from AEC, CC, and CD also have ideal points that are not significantly different from zero. Figure~\ref{densidades} presents the posterior marginal distributions of the ideal points for four selected senators: Alexander López Maya (PDA, \( \beta_2 \)), Cecilia López Montaño (PL, \( \beta_{22} \)), Mario Uribe Escobar (CD, \( \beta_{110} \)), and Zulema del Carmen Jattin Corrales (PU, \( \beta_{142} \)). The latter two were implicated in the parapolitics scandal. Their marginal distributions are notably asymmetric, reflecting greater uncertainty in the estimation of their ideal points. This increased uncertainty is likely due to limited participation in roll call votes, resulting from legal proceedings that affected their ability to engage in legislative activity.

The senators most likely positioned at the extremes of the ideological spectrum are primarily affiliated with either the opposition or the governing coalition. On the left, the seven senators most likely to occupy extreme positions all belong to the PDA. On the right, four of the five most right-leaning senators are members of the PU. Those with the highest posterior probability of having an ideal point less than \(-1\) include Jorge Enrique Robledo Castillo (PDA, 99.6\%), Jaime Dussan Calderón (PDA, 99.3\%), Alexander López Maya (PDA, 98.8\%), Luis Carlos Avellaneda Tarazona (PDA, 97.8\%), Gloria Inés Ramírez Ríos (PDA, 95.2\%), and Gustavo Francisco Petro Urrego (PDA, 94.8\%). On the opposite end, the senators most likely to have an ideal point greater than 1 are María Isabel Mejía Marulanda (PU, 76.9\%), Jorge Enrique Gómez Montealegre (CV, 71.3\%), Manuel Guillermo Mora Jaramillo (PU, 70.9\%), Martha Lucía Ramírez de Rincón (PU, 68.7\%), and Dilian Francisca Toro Torres (PU, 62.2\%).

\begin{figure}[!b]
    \centering
    \includegraphics[scale=0.8]{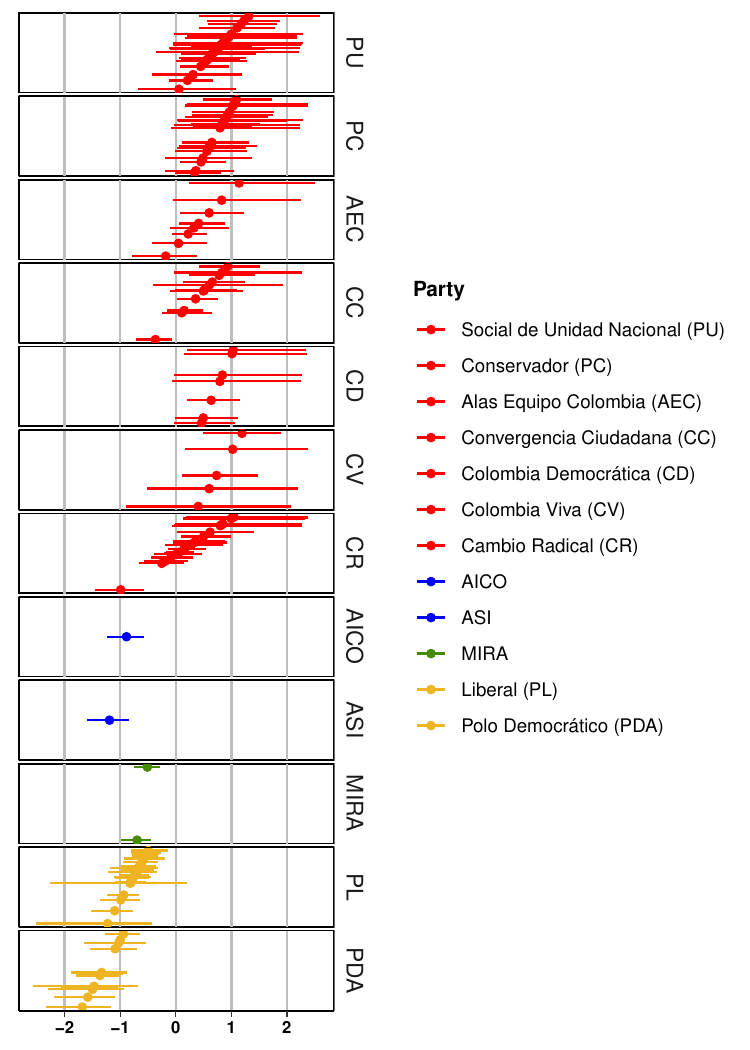}
    \caption{Estimated ideal points by party based on anchor legislators positioned at opposite ends of the spectrum. Points indicate posterior means, and horizontal lines represent symmetric 95\% credibility intervals computed from posterior percentiles.}
    \label{fig_Credibilidad_partidos}
\end{figure}

For the PL party, the probability of any senator having an ideal point less than \(-1\) does not exceed 68\%. Among its 22 members, 20 have a probability below 35\% of being positioned on the far left (91\%), in stark contrast to the PDA, where 9 out of 12 senators (75\%) have a probability greater than 80\% of having an ideal point below \(-1\). Within the coalition, the PU and PC exhibit the highest probabilities of containing senators with ideal points greater than 1. Conversely, CR shows the lowest probability of being positioned on the far right, with 15 out of 23 senators (65.2\%) having less than a 3\% chance of reaching an ideal point above 1. Among minor parties and independents, Jesús Enrique Piñacué Achicué (ASI) stands out with an 84.9\% probability of having an ideal point below \(-1\). For other senators in these groups, the probability of being located at either ideological extreme is generally low.

Several members of the governing coalition have the highest probability of being located near the center of the political space. Among the 16 senators with a probability greater than 40\% of having an ideal point between \(-0.2\) and \(0.2\), 9 belong to CR (56.3\%) and 3 to AEC (18.8\%). The senators most likely to fall within this central range include Juan Carlos Restrepo Escobar (CR, 85.7\%), Antonio del Cristo Guerra de la Espriella (CR, 84\%), Juan Carlos Martínez Sinisterra (CC, 66.8\%), Jorge Enrique Vélez García (CR, 64.7\%), and Rodrigo Lara Restrepo (CR, 63.3\%). In contrast, most senators from the CD, PU, and PC have less than a 20\% chance of being located in the center. Senators from minority parties and independents have virtually no probability of occupying this position.

Figure \ref{fig_Credibilidad_partidos} reveals two broad ideological groupings: One to the left of zero and one to the right. The left-leaning group includes the opposition, independents, and minority parties, while the right-leaning group consists of the governing coalition. Within the coalition, two distinct subgroups emerge: CC, CD, CV, PC, and PU are positioned farther to the right, whereas AEC and CR have many ideal points concentrated near the center. The proximity of some AEC and CR senators to those of PL reflects their shared political origins. Among the opposition, PDA is situated further to the left than PL, with most of its members located left of \(-1\), while the majority of PL senators fall to the right of that threshold. Minority parties cluster around \(-1\), and independents occupy a region similar to that of PL, indicating ideological proximity. 

The empirical evidence suggests that the latent trait underlying nominal voting in the Colombian Senate (2006–2010), as captured by the Euclidean model, is not ideological in the traditional left–right sense. For example, the PL, historically opposed to the PC, has not been considered a leftist party in the last three decades \citep{uribe2020partidos}, yet it appears aligned with PDA in the ideological space. Similarly, MIRA, a center-right party, is positioned near PL on the left, reinforcing this interpretation. A government-versus-opposition interpretation is also inadequate, as independents and minority parties did not consistently align with either bloc. Instead, the latent dimension appears to capture a broader contrast between opposition and non-opposition forces, shaped by the structural imbalance between the relatively small opposition, minority, and independent groups and the dominant governing coalition. This pattern of non-ideological polarization has been observed in Colombia and other Latin American legislatures when analyzing roll call behavior \citep{londono2014efectos, luque2022bayesian}.

\begin{figure}[!htb]
    \centering
    \includegraphics[scale=0.6]{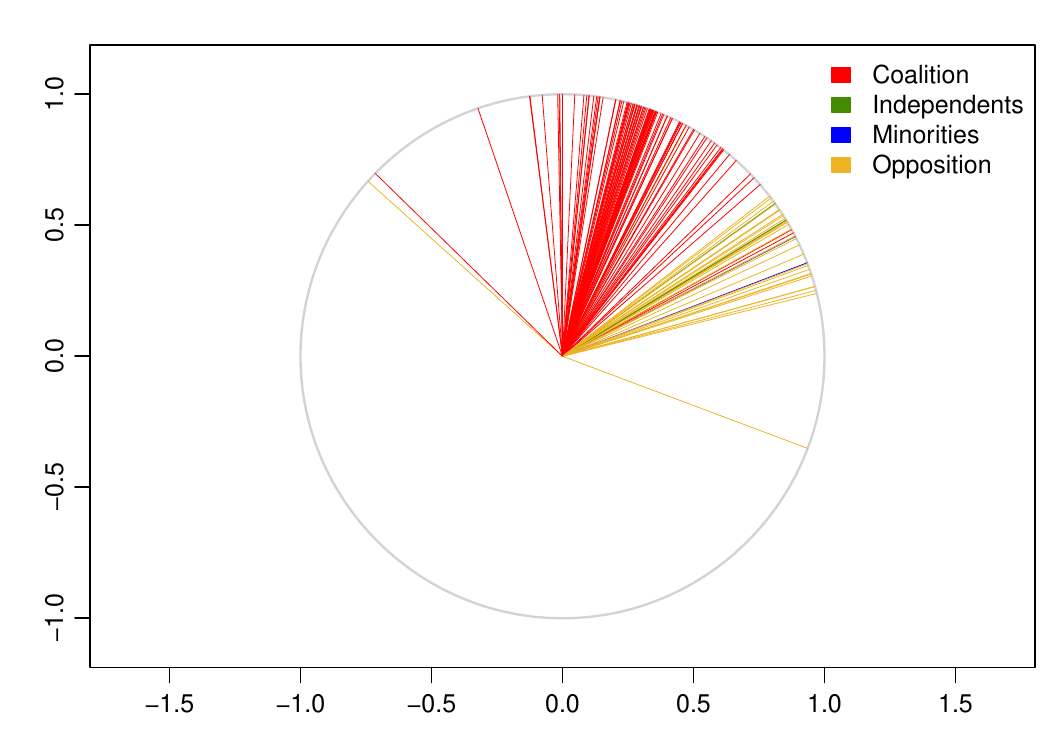}
    \caption{Posterior mean ideal points on the unit circle.}
    \label{fig_ideal_points_circle}
\end{figure}

Figure \ref{fig_ideal_points_circle} displays the posterior means of ideal points estimated under the unit circle model, providing a valuable complement to the Euclidean representation. While the Euclidean model assumes a linear ideological continuum (typically interpreted as left versus right), the spherical representation embeds ideal points on a circular manifold. This framework is particularly suited to contexts where ideological extremes may converge in behavior, or where political alignments defy simple one-dimensional interpretation. The circular map preserves key patterns identified in the Euclidean model: Most notably, the arc-based clustering of opposition, independents, and minorities opposite the governing coalition. 

\begin{figure}[!htb]
    \centering
    \includegraphics[scale=0.21]{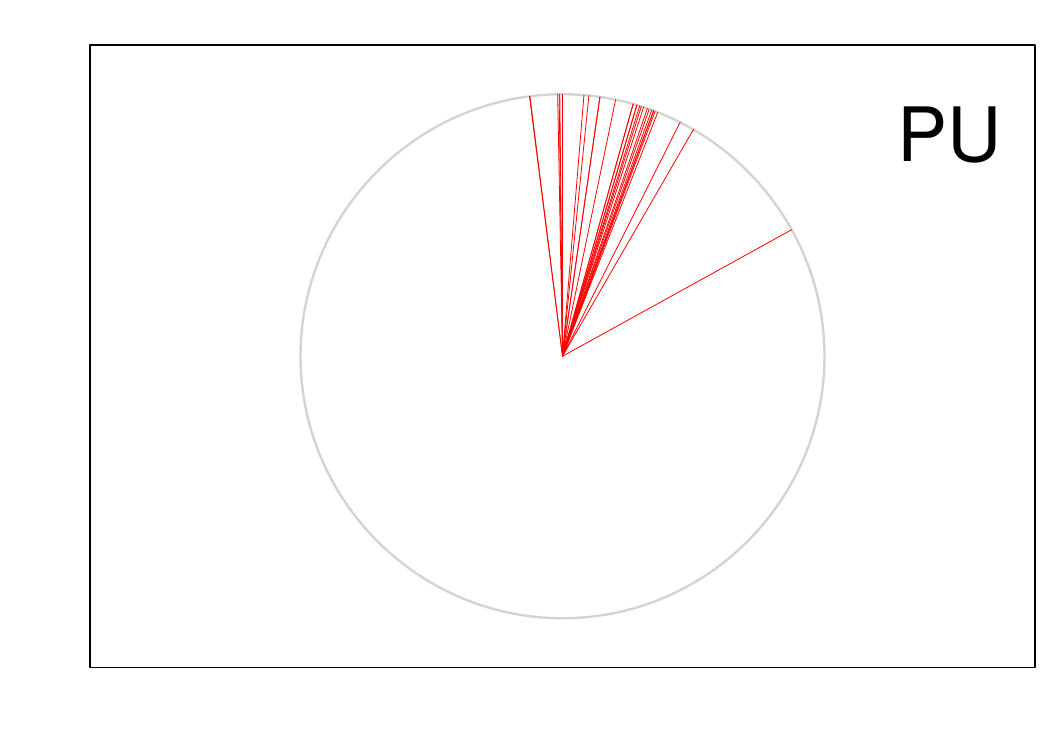}
    \includegraphics[scale=0.21]{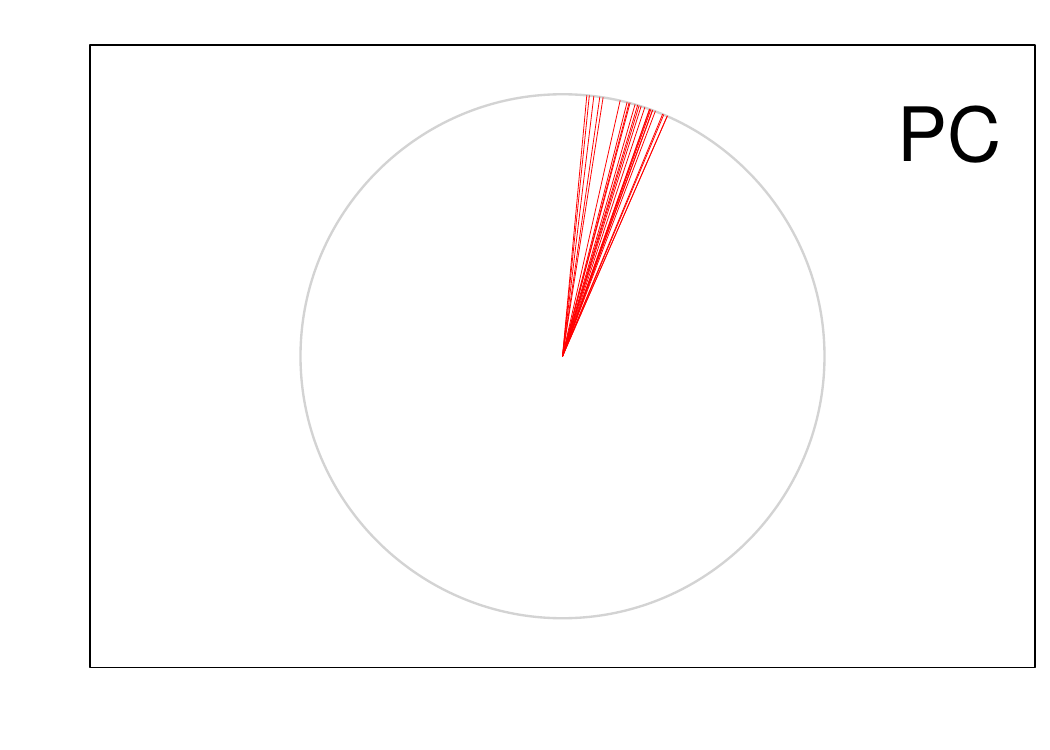}
    \includegraphics[scale=0.21]{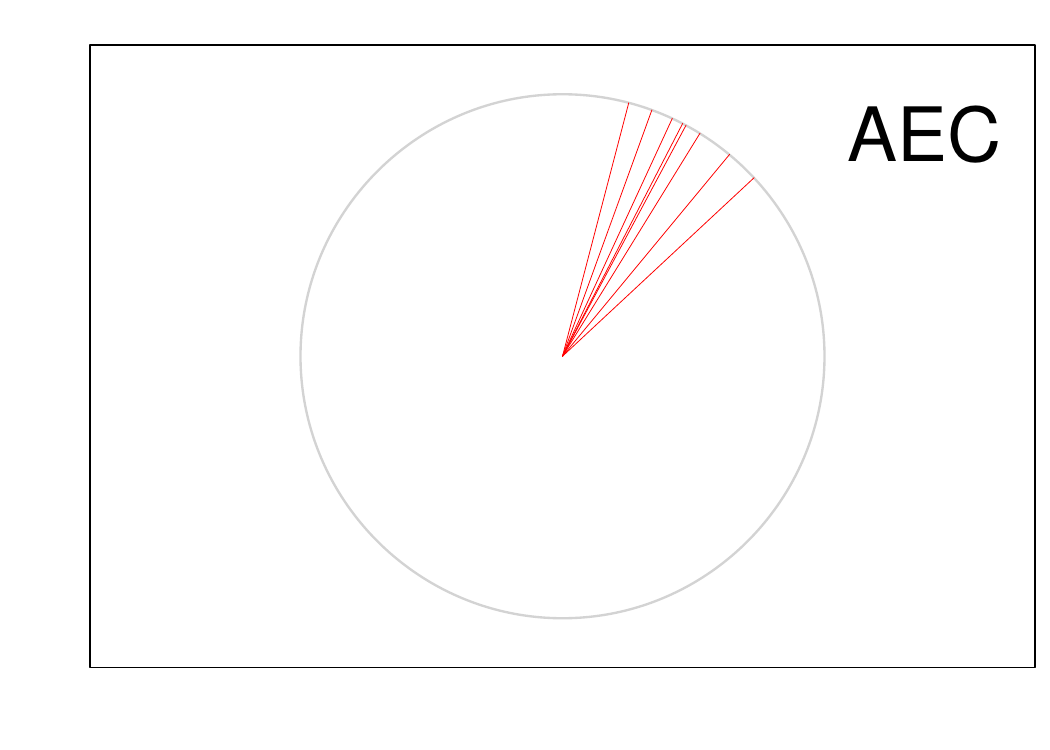}
    \includegraphics[scale=0.21]{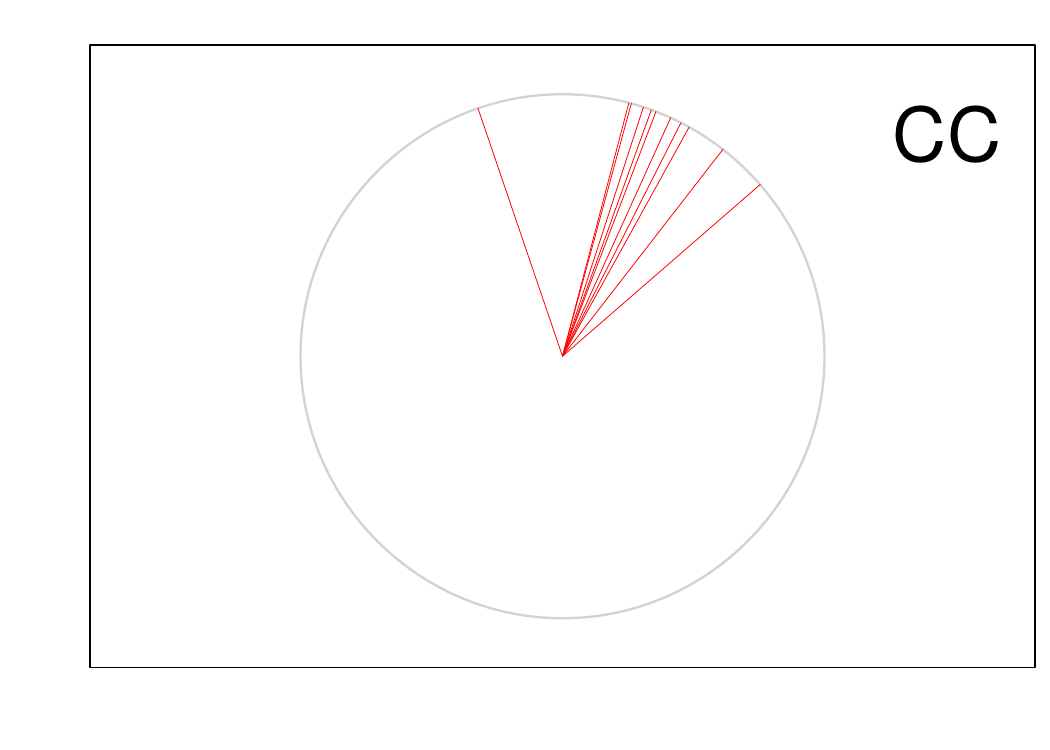} \\
    \includegraphics[scale=0.21]{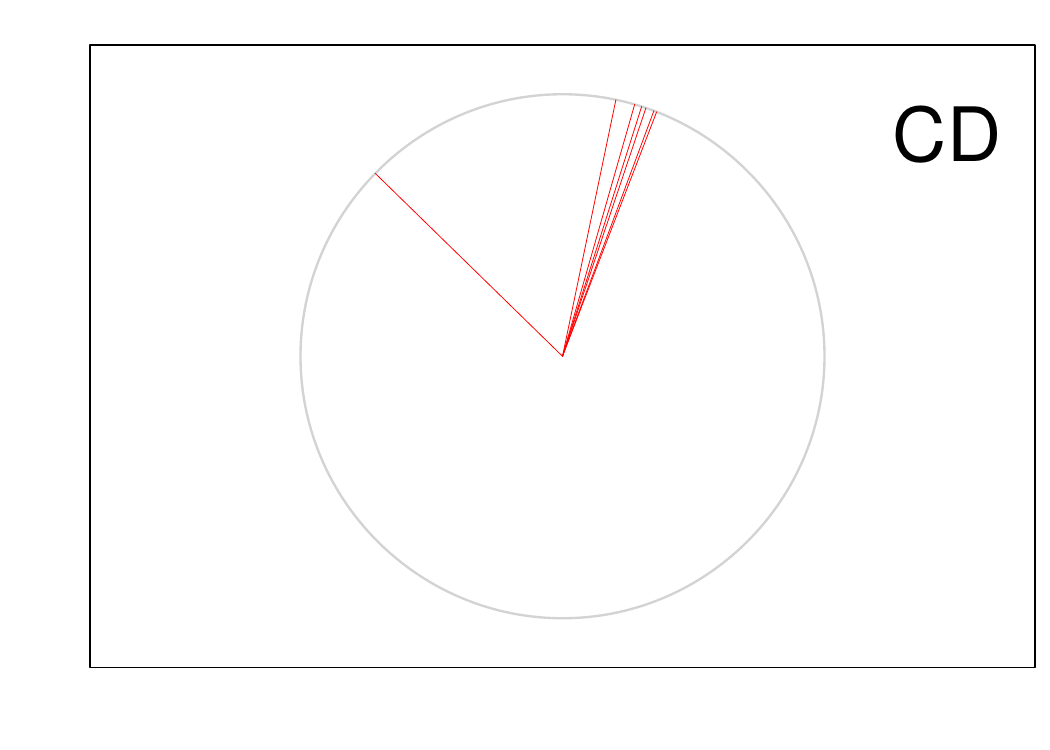} 
    \includegraphics[scale=0.21]{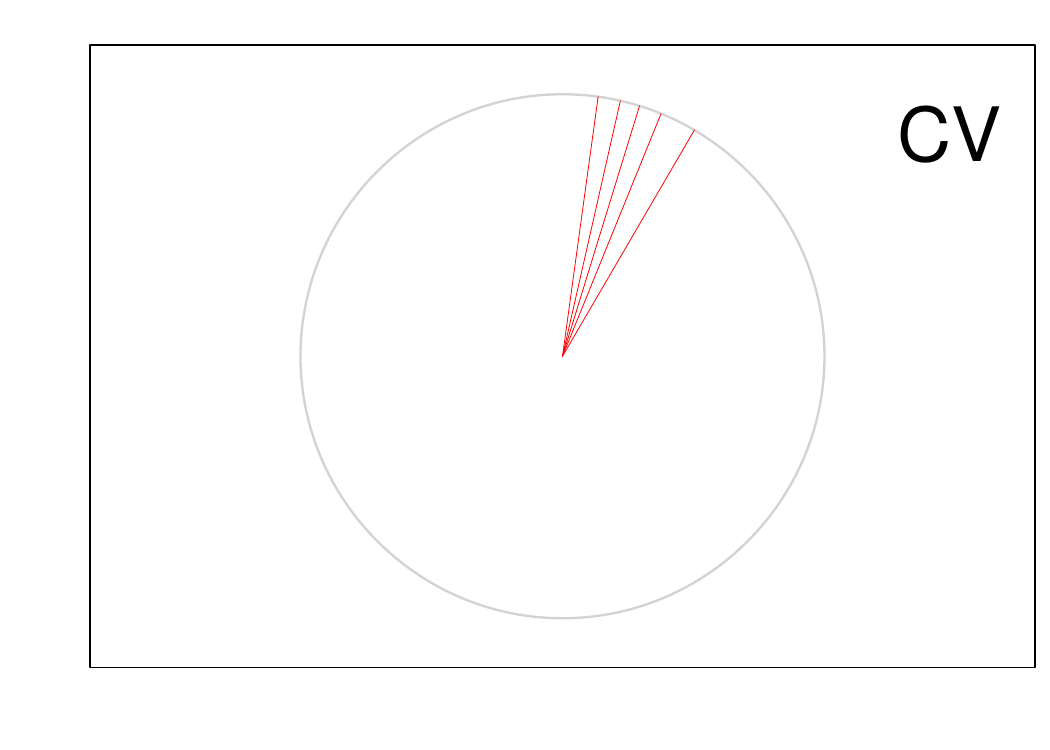}
    \includegraphics[scale=0.21]{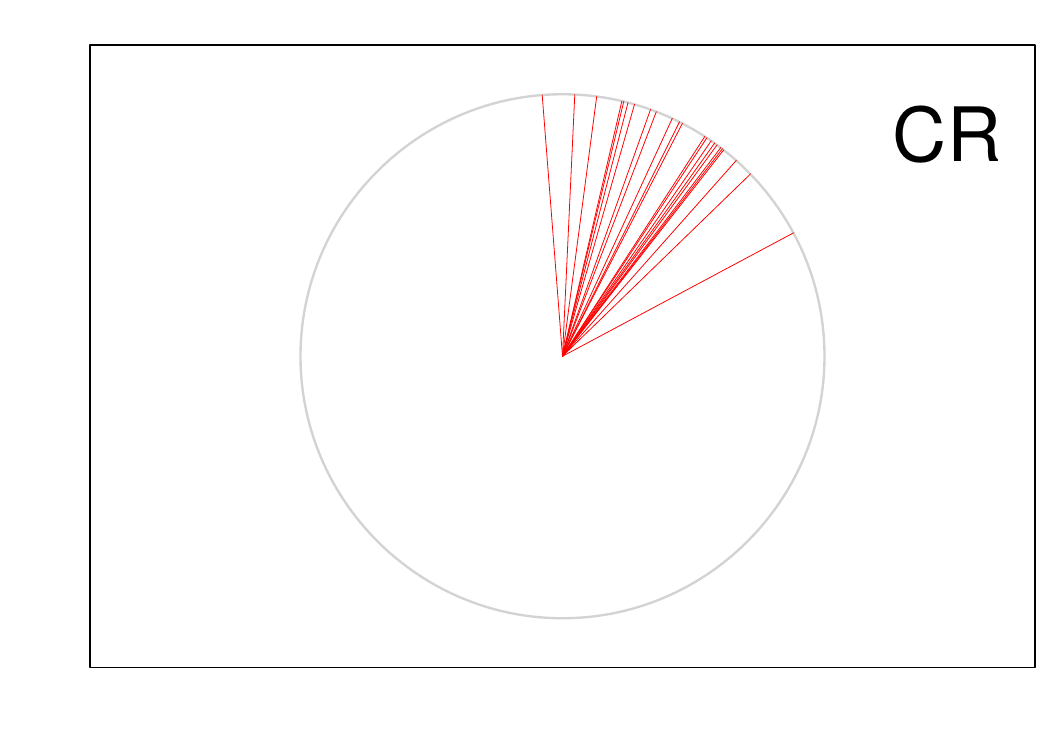}
    \includegraphics[scale=0.21]{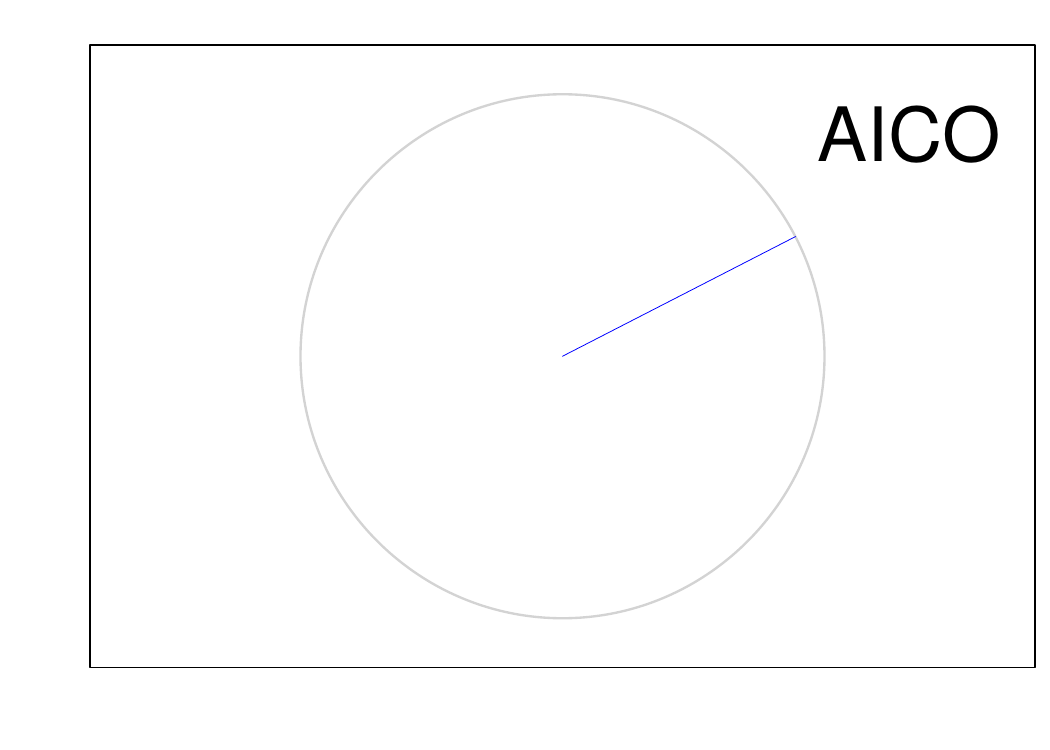} \\
    \includegraphics[scale=0.21]{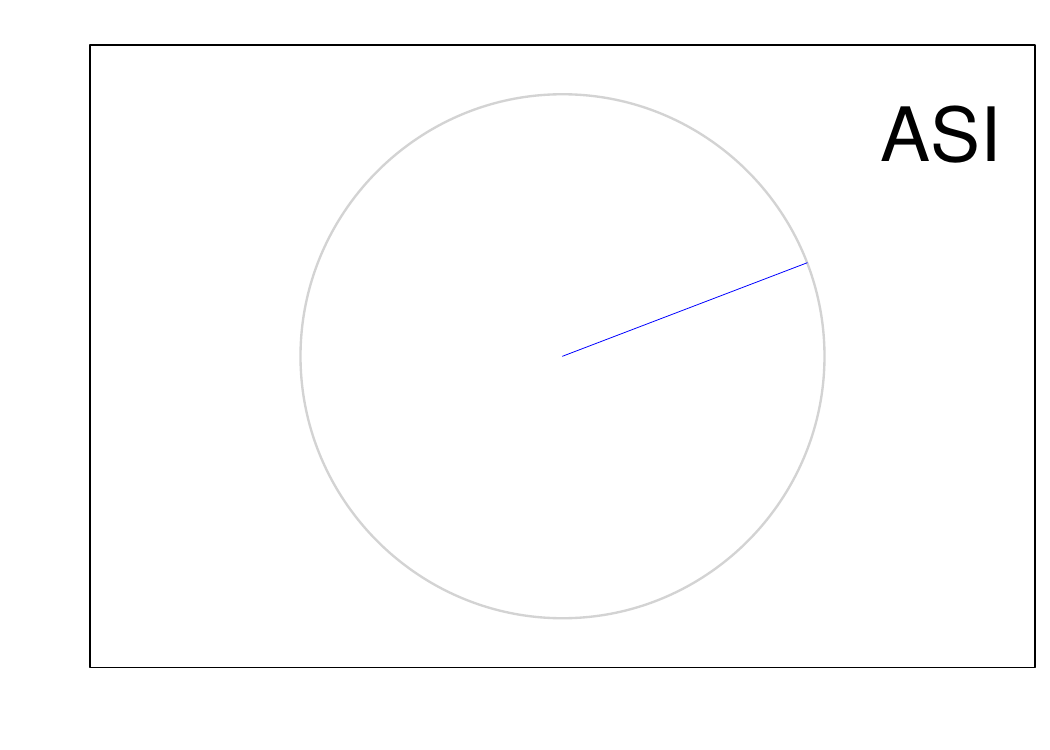}
    \includegraphics[scale=0.21]{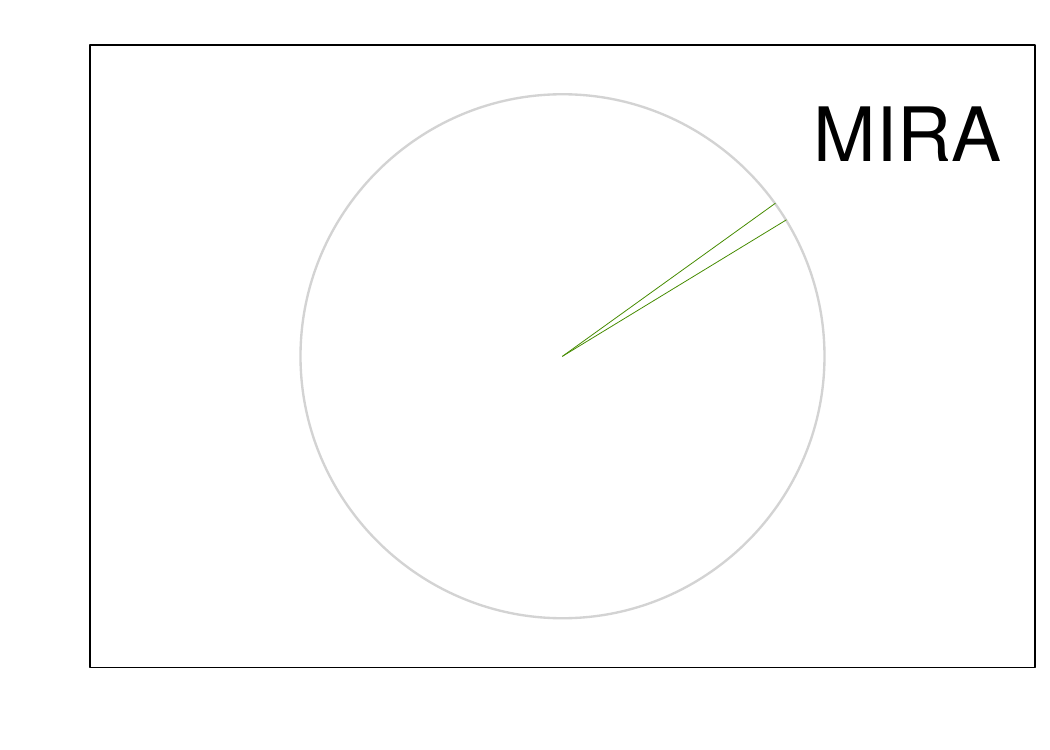}
    \includegraphics[scale=0.21]{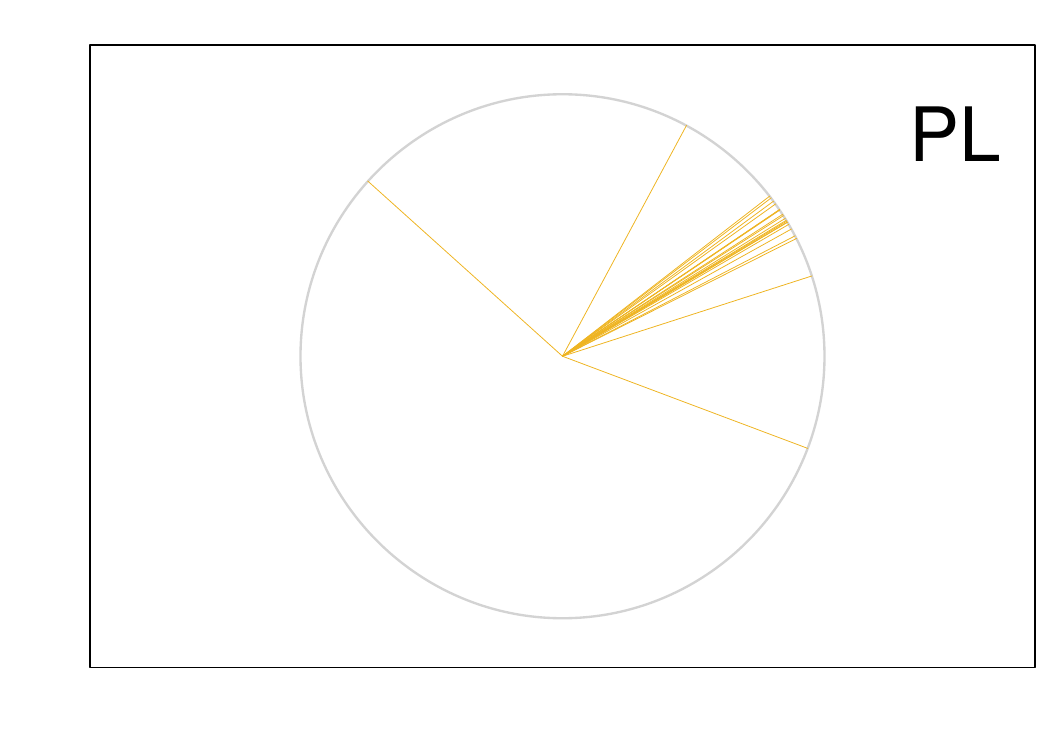}
    \includegraphics[scale=0.21]{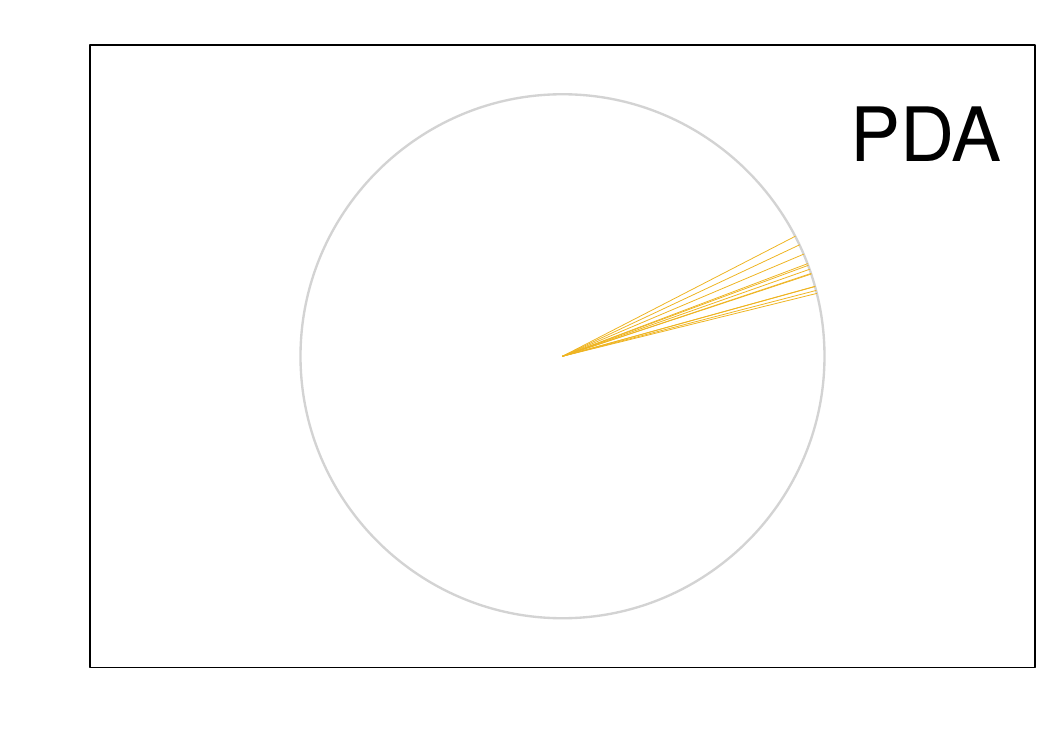}
    \caption{Posterior mean ideal points on the unit circle by political group.}
    \label{fig_ideal_points_circle_parties}
\end{figure}

Figure~\ref{fig_ideal_points_circle_parties} presents the posterior means of ideal points under the unit circle model, disaggregated by political party. This disaggregation highlights several important patterns. For example, the PL, historically opposed to the PC, appears near the PDA in the Euclidean space, despite their divergent ideological trajectories. Likewise, MIRA, typically classified as a center-right party, is located close to left-leaning groups. The circular model attenuates these inconsistencies by decoupling ideological similarity from linear spatial assumptions, allowing proximity to reflect strategic alignment or voting behavior rather than strict ideological equivalence.

However, the spherical layout reframes groupings by eliminating the directional bias inherent in linear projections. With no privileged center and all directions treated symmetrically, it provides a more neutral view of political dispersion and cohesion. For instance, CR and AEC appear near the opposition cluster, echoing their ambiguous alignment in the Euclidean map and the fact that many of their ideal points are indistinguishable from zero. In contrast, PU, PC, and CV are concentrated along a denser segment of the coalition arc, reinforcing their roles as more cohesive or ideologically aligned members of the governing bloc. These patterns further support the interpretation that the latent trait captured by the model reflects not a conventional left–right spectrum, but rather a broader opposition–non-opposition divide.

Importantly, the circular projection clarifies the asymmetry in political behavior observed earlier. The opposition–minority–independent segment appears tightly grouped, indicating higher behavioral cohesion, while the coalition stretches across a broader arc. This visual asymmetry aligns with the structural imbalance noted in the Euclidean analysis and reinforces the interpretation of the latent dimension as a strategic alignment spectrum rather than an ideological one. 

\begin{figure}[!htb]
    \centering
    \includegraphics[scale=0.6]{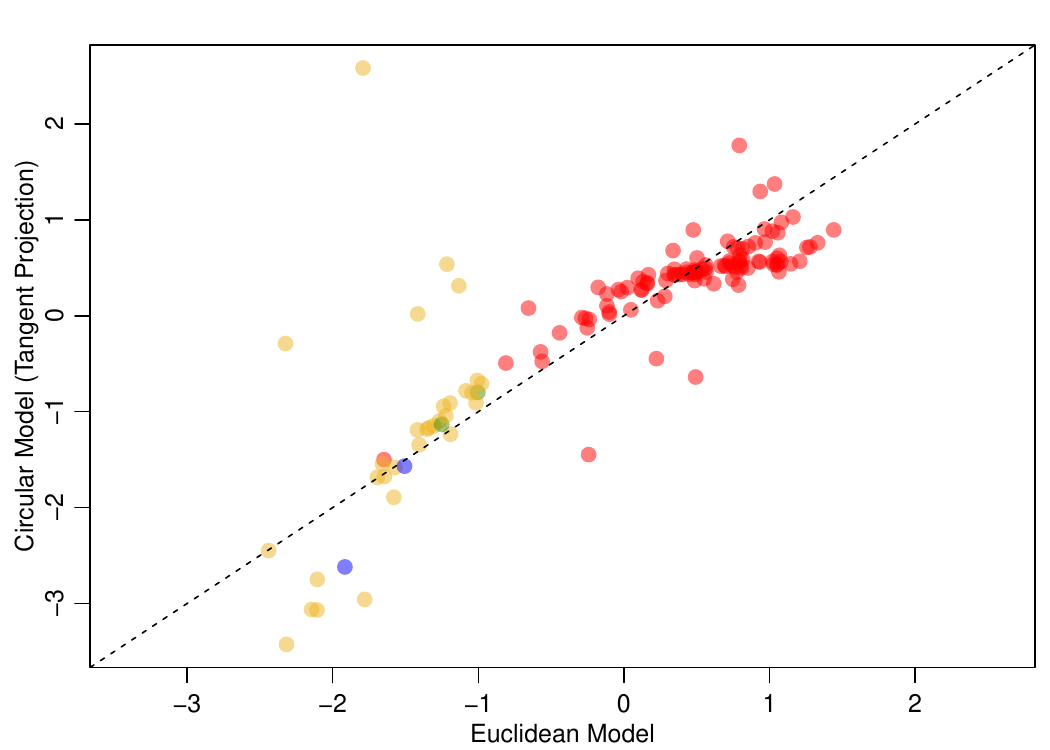}
    \caption{Comparison of posterior means from the Euclidean model and tangent space projections from the circular model. The dashed line indicates \( x = y \).}
    \label{fig_ideal_points_circular_vs_euclidean}
\end{figure}

To compare the results from both model specifications, we examine the posterior means of the estimated ideal points obtained under the Euclidean model and those projected onto the tangent space from the circular model. Three senators were excluded from the analysis due to extreme values in the tangent space, which arise when their posterior mean angles in the circular model are close to \( \pi \), and thus yield very large values after applying the log map centered at \( \pi/2 \). As shown in Figure~\ref{fig_ideal_points_circular_vs_euclidean}, the majority of points lie close to the identity line \( x = y \), suggesting strong agreement between the two models. The estimated correlation between both sets of ideal points is 0.824, providing further reassurance about the robustness of the latent trait captured in both geometric representations.

Nevertheless, a few senators deviate noticeably from the identity line, indicating meaningful differences in their estimated positions depending on the underlying geometry. In particular, changing scale and geometry most affects the ideal points of extremist opposition legislators. These discrepancies highlight the distinct assumptions of each model: While the Euclidean framework imposes a linear structure that may misplace ideologically atypical or strategically ambiguous legislators, the circular model permits more flexible configurations. Such flexibility can better accommodate actors whose voting behavior diverges from the dominant left–right or coalition–opposition dichotomies. Hence, this comparison not only supports the consistency of the main findings but also underlines the added interpretive value of the spherical approach in capturing political complexity within a fragmented and evolving party system.

\subsection{Association between parapolitics and ideal points}

\begin{figure}[!htb]
    \centering
    \includegraphics{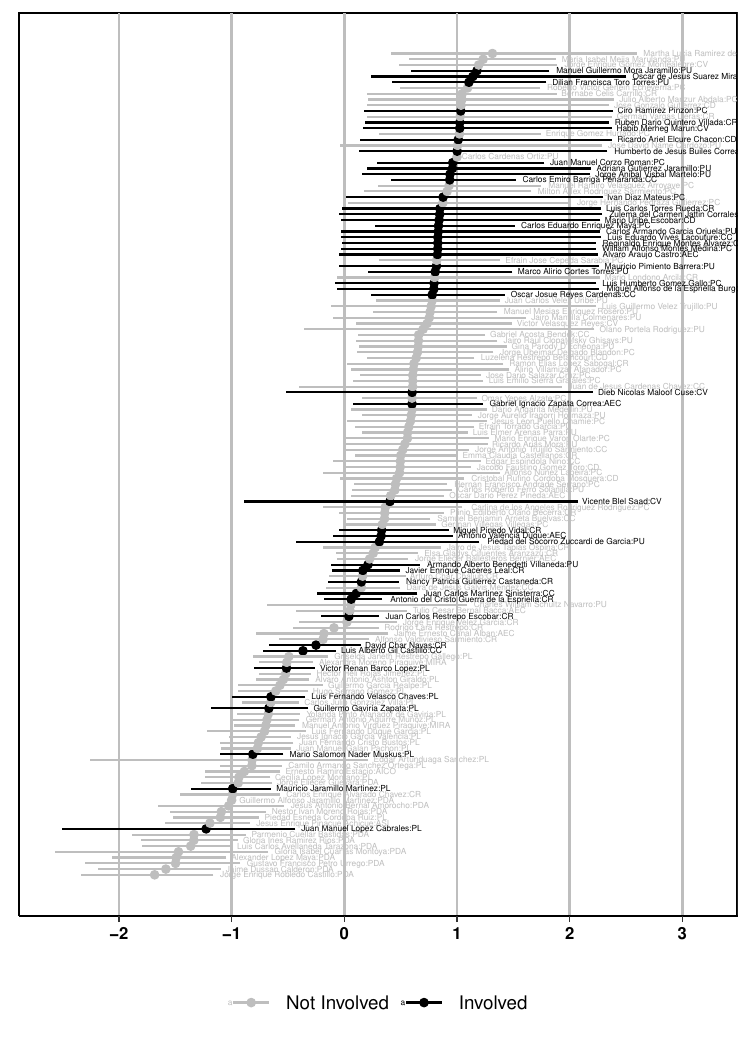}
    \caption{Classification of Euclidean ideal points based on whether senators were involved in the parapolitics scandal.}
    \label{credibilidadParapolitica}
\end{figure}

\begin{figure}[!htb]
    \centering
    \includegraphics[scale=0.6]{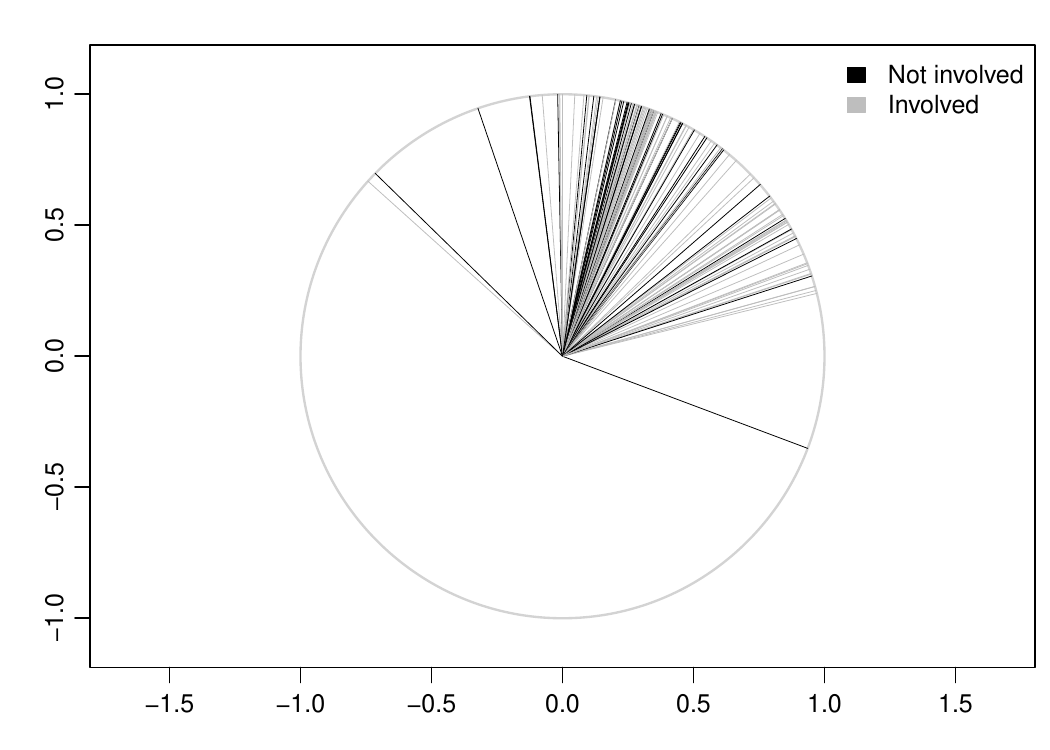}
    \caption{Classification of ciruclar ideal points based on whether senators were involved in the parapolitics scandal.}
    \label{fig_ideal_points_circle_parapolitics}
\end{figure}

Figures \ref{credibilidadParapolitica} and \ref{fig_ideal_points_circle_parapolitics} present a comparative view of senators' estimated ideal points based on their involvement in the parapolitics scandal, using both Euclidean and spherical representations. In the Euclidean model, a significant majority (83\%) of implicated senators have ideal points to the right of zero, with over half (57.4\%) concentrated in the narrow interval from 0.78 to 1.17. This clustering suggests a pattern of alignment with the governing coalition. Moreover, their credibility intervals tend to be wide, reflecting high uncertainty stemming from low participation in legislative votes, often due to legal proceedings. 

The circular representation in Figure \ref{fig_ideal_points_circle_parapolitics} reinforces this observation, showing that implicated senators are mostly confined to a specific arc of the circle, corresponding roughly to the right-hand side of the Euclidean spectrum. This consistency across geometries suggests that while the latent trait is not strictly ideological, those involved in the scandal were largely aligned with the dominant bloc. At the same time, the spherical model nuances this alignment by showing dispersion around the arc, capturing subtle heterogeneity in behavior that may be missed in a linear projection. Overall, both figures illustrate that parapolitics involvement is associated with a tendency to cluster in the non-oppositional region of the latent space, supporting the broader interpretation that the underlying dimension captures alignment versus dissent rather than traditional ideological divides.

To assess the relationship between senators' revealed policy preferences and their involvement in the parapolitics scandal, we fit a logistic regression model using ideal points as predictors. This was done for both the Euclidean and circular models of ideal point estimation. In the Euclidean case, we used posterior draws of the ideal points obtained from the standard latent variable model. For the circular model, we projected the posterior samples onto the tangent space at a fixed reference point, ensuring comparability across geometries. Importantly, we removed two anchor legislators whose constrained positions could bias the results, and for the circular model, we further excluded posterior samples with extreme tangent values to stabilize inference.

We implemented this analysis by fitting the logistic regression model separately for each posterior draw of the ideal points. This approach allows us to propagate uncertainty from the latent space estimation into the logistic regression, producing a posterior distribution over the regression coefficients and the area under the ROC curve (AUC). The logistic model is given by $\textsf{logit}(y_i) = \beta_0 + \beta_1 x_i$, where $y_i$ denotes whether senator $i$ was involved in the parapolitics scandal, and $x_i$ is their estimated ideal point.

Results are provided in Table \ref{tab_logistic_regression_results}. For the Euclidean model, the posterior mean of the slope parameter was 0.52, with a 95\% credible interval of (0.32, 0.73), and a posterior AUC of 0.62. The circular model yielded nearly identical results, with a slope of 0.53 (95\% CI: 0.23, 0.85) and an AUC of 0.61. In both cases, the posterior distributions of $\beta_1$ are bounded away from zero with high probability, providing strong evidence that senators with more extreme or polarized ideal points, particularly on the right side of the spectrum, were significantly more likely to be involved in the parapolitics scandal. These findings are consistent across both geometric representations, underscoring the robustness of the relationship.

\begin{table}[!htb]
\centering
\begin{tabular}{lcccccc}
\hline
 & \multicolumn{3}{c}{Euclidean Model} & \multicolumn{3}{c}{Circular Model} \\
\cline{2-4} \cline{5-7}
Statistic & Intercept & Slope & AUC & Intercept & Slope & AUC \\
\hline
Mean & -2.317 & 0.522 & 0.624 & -2.330 & 0.530 & 0.611 \\
\hline
Q 2.5\%  & -2.993 & 0.320 & 0.571 & -3.371 & 0.231 & 0.542 \\
Q 97.5\% & -1.681 & 0.734 & 0.675 & -1.394 & 0.852 & 0.671 \\
\hline
Q 0.5\%  & -3.249 & 0.262 & 0.555 & -3.629 & 0.168 & 0.518 \\
Q 99.5\% & -1.501 & 0.813 & 0.691 & -1.199 & 0.933 & 0.685 \\
\hline
\end{tabular}
\caption{Posterior means and credible intervals (95\% and 99\%) for the logistic regression coefficients and AUC, comparing Euclidean and Circular ideal point models.}
\label{tab_logistic_regression_results}
\end{table}

The moderate AUC values suggest that while ideal points are not perfect predictors, they still provide meaningful discriminatory power, distinguishing involved from uninvolved senators better than chance. This is particularly notable given that the model relies on a single latent explanatory variable derived from legislative behavior alone. It highlights how voting records, when summarized into ideological dimensions, can reveal deeper structural alignments between policy positioning and extralegal political dynamics.

These results have important implications. They suggest that involvement in the parapolitics scandal was not randomly distributed across the ideological spectrum but instead concentrated among legislators with specific voting patterns. This supports the hypothesis that parapolitics was entangled with broader patterns of legislative alignment and coalition formation, particularly among right-leaning and pro-government factions \citep{avila2012politics}. The consistency between the Euclidean and circular results further validates the conclusion and demonstrates that our methodological approach is resilient to the choice of latent geometry.

\section{Discussion}

This study comprehensively analyzes roll call voting behavior in the Colombian Senate during the 2006–2010 legislative period using two distinct latent factor models: A traditional Euclidean spatial voting model and its circular extension. We began by consolidating a detailed database of 147 senators and 136 plenary roll call votes, integrating political affiliation, attendance, abstention, and parapolitics involvement. Using this data, we estimated senators’ latent policy preferences (ideal points) under both geometric frameworks and analyzed their association with the para-politics scandal, one of the most salient political events of the period.

We first applied a one-dimensional Euclidean model, fixing anchor legislators to resolve identifiability and relying on standard MCMC methods for inference \citep{luque2023bayesian}. In parallel, we implemented the spherical latent factor model of \citet{yu2020spherical}, embedding ideal points on the unit circle. We employed a tangent space projection of the circular estimates to compare models and aligned them to a common reference frame. Across both approaches, we observed similar patterns in the posterior means of ideal points. We noted a strong correlation (0.824) between them, underscoring the robustness of the underlying latent dimension across geometries.

Our findings indicate that this latent dimension is not best interpreted as a traditional left–right ideological continuum. Instead, both models reveal an opposition–non-opposition structure, where independent, minority, and opposition parties cluster in contrast to the governing coalition. This aligns with prior research on Latin American legislatures that emphasizes the fluidity of ideological labels and the role of strategic alignment in shaping voting behavior \citep{londono2014efectos, luque2023bayesian, luque2024operationalizing}.

One of the central contributions of this paper is the examination of how ideal point estimates relate to involvement in the para-politics scandal. Based on a thorough review of historical records and media reports, we categorized senators as either involved or uninvolved in the scandal. Using logistic regression models fitted across posterior samples of ideal points, we quantified the relationship between voting behavior and para-politics involvement. In both the Euclidean and circular models, the slope parameter was consistently positive and statistically distinct from zero, suggesting that senators aligned with the dominant coalition and, thus, occupying positions farther to the right or along specific arcs, were significantly more likely to be implicated. The AUC values, though moderate (around 0.611 and 0.624), demonstrate that the latent preferences derived from roll call data possess discriminatory power in classifying scandal involvement, even when used as the sole predictor.

These findings imply that the para-politics scandal was not randomly distributed across the political spectrum but was instead entangled with deeper structural patterns of alignment and influence within the Senate. The use of two modeling approaches, yielding consistent conclusions, reinforces the empirical strength of this result and highlights the methodological flexibility afforded by the circular model when accounting for ambiguous or extreme voting patterns.

Nevertheless, several limitations remain. First, our analysis focuses exclusively on the Senate during a single legislative term. Although this period coincides with the height of the parapolitics scandal, the origins of many political and paramilitary alliances can be traced back to earlier legislative cycles \cite{avila2012politics,londono2014efectos,bakiner2020endogenous}. Second, while the circular model accommodates greater behavioral heterogeneity, its interpretability may be less intuitive for readers accustomed to linear ideological spectra. Lastly, our treatment of missing data differs across models (removed in the Euclidean case and imputed in the circular model). However, we demonstrate that both strategies yield consistent results, in line with the simulation findings of \citet{luque2023bayesian}.

Future work could extend this analysis along several dimensions \citep{yu2020spherical,moser2021multiple}. A natural next step is to estimate ideal points for the Colombian House of Representatives during the same period or to expand the temporal scope to cover both chambers from 2002 to 2010. Such an approach would offer a deeper understanding of how paramilitary influence evolved over time and across institutions. A hierarchical Bayesian framework could be developed to jointly estimate ideal points and logistic regression parameters, allowing the model to account for uncertainty propagation in a unified structure \citep{gelman2013bayesian}. Finally, further refinements could incorporate covariates such as party membership, electoral district, or legislative seniority, providing a richer account of the factors that shaped political alignment during this pivotal period in Colombian history.

\section*{Statements and Declarations}

The authors declare that they have no known competing financial interests or personal relationships that could have appeared to influence the work reported in this article.

During the preparation of this work the authors used ChatGPT-4-turbo in order to improve language and readability. After using this tool, the authors reviewed and edited the content as needed and take full responsibility for the content of the publication.

\bibliographystyle{apalike}
\bibliography{references_v2.bib}

\end{document}